\documentclass[a4paper,12pt]{article}
\usepackage{a4wide,latexsym}
\usepackage{psfig,graphicx}

\newcounter{fig2}

\renewcommand{\theequation}{\arabic{section}.\arabic{equation}}
\renewcommand{\thefigure}{\arabic{figure}\alph{fig2}}

\newcommand{\bea}{\begin{eqnarray}}
\newcommand{\eea}{\end{eqnarray}}
\newcommand{\be}{\begin{equation}}
\newcommand{\ee}{\end{equation}}
\newcommand{\pkt}{\; .}
\newcommand{\kma}{\; ,}
\newcommand{\eqn}[1]{(\ref{#1})}

\newcommand{\intki}{\int\!\frac{{d^3}k }{(2\pi)^32\omega_{0i}}\,}

\newcommand{\calm}{{\cal M}}

\newcommand{\calv}{{\cal V}}
\newcommand{\calg}{{\cal G}}
\newcommand{\calp}{{\wp}}

\newcommand{\calf}{{\cal F}}
\newcommand{\calc}{{\cal C}}

\newcommand{\cale}{{\cal E}}
\newcommand{\call}{{\cal L}}
\newcommand{\bfk}{{\bf k}}
\newcommand{\bfx}{{\bf x}}

\newcommand*{\del}{\partial}

\begin{document}

\begin{titlepage}
\begin{flushright}
DO-TH-01/11\\
September 2001\\
revised November 20, 2001
\end{flushright}

\vspace{20mm}
\begin{center}
{\Large \bf Nonequilibrium evolution in scalar $O(N)$ models 
  with spontaneous symmetry breaking  }

\vspace{10mm}

{\large  J\"urgen Baacke \footnote{
e-mail:~\texttt{baacke@physik.uni-dortmund.de}}} and {\large Stefan Michalski
\footnote{e-mail:~\texttt{stefan.michalski@uni-dortmund.de}} } \\
{\large Institut f\"ur Physik, Universit\"at Dortmund} \\
{D - 44221 Dortmund, Germany}
\\
\vspace{8mm}
\end{center}

\begin{abstract}
\noindent
We consider the out-of-equilibrium evolution of a classical condensate field and its
quantum fluctuations for a scalar $O(N)$ model with spontaneously
broken symmetry.
In contrast to previous studies we do not consider the large $N$ limit,
but the case of finite $N$, including $N=1$, i.e., plain $\lambda \phi^ 4$
theory.  The instabilities encountered in the one-loop approximation
are prevented, as in the large-$N$ limit, by  back reaction of the
fluctuations on themselves, or,  equivalently, by including a resummation of bubble diagrams.
 For this resummation and its renormalization we use formulations developed recently
based on the effective action formalism
of Cornwall, Jackiw and Tomboulis.
The formulation of renormalized equations for finite $N$ derived here
represents a useful tool for simulations with realistic models.
Here we concentrate on the phase structure of such models.
We observe the transition between the spontaneously
broken and the symmetric phase at low and high energy densities, respectively.
This shows that the typical structures expected in thermal equilibrium are encountered
in nonequilibrium dynamics even at early times, i.e.,
before an efficient rescattering can lead to thermalization.
\end{abstract}

\end{titlepage}

%%%%%%%%%%%%%%%%%%%%%%%%%%%%%%%%%%%%%%%%%%%%%%%%%%%%%%%%%%%%%%%%%%%%%%%%%%%%%%%%%%%

\section{\label{intro}Introduction}
\setcounter{equation}{0}

The investigation of the $O(N)$ vector model at large $N$ has a
long-standing history in quantum field theory
\cite{Coleman:1974jh,Dolan:1974qd,BarMosh}.
The dynamical exploration of  nonequilibrium properties
of such models has been developed only recently
\cite{Calzetta:1987ey,Cooper:1994hr,Boyanovsky:1994xf,Boyanovsky:1995me,
Boyanovsky:1995em,Baacke:1997se}.
The nonequilibrium aspects of spontaneous symmetry breaking in
particular have been studied
in the large-$N$ approximation in Refs. \cite{Cooper:1997ii,Boyanovsky:1996sq,
Boyanovsky:1998zg,Boyanovsky:1999yp,Baacke:2000fw}.

At finite $N$ there have been several approaches to formulating a Hartree-type
interaction between the quantum modes.
These lead, in general, to problems with renormalization
\cite{Lenaghan:2000si,DM}.
On the other hand, without such a back reaction the nonequilibrium systems with spontaneous
symmetry breaking run into seemingly unphysical instabilities
(see e.g. Fig. 9 in \cite{Baacke:2000fw} for an illustration).
For equilibrium quantum field theory at finite temperature this interaction between
quantum modes is taken into account by resummation of bubble  diagrams, i.e.,
by including daisy and super-daisy diagrams. These are essential
for studying the phase transitions between the spontaneously broken
phase and the phase with restored symmetry.  Techniques of bubble
resummation have been developed \cite{Amelino-Camelia:1993nc,
Amelino-Camelia:1997dd,ChiHa}
based on the effective action formalism of Cornwall, Jackiw and Tomboulis (CJT)
\cite{Cornwall:1974vz}.
Recently there have been some new resummation schemes
\cite{Nemoto:2000qf,Verschelde:2001dz,Verschelde:2000ta,Smet:2001un}
 which can be consistently renormalized. While Ref. \cite{Nemoto:2000qf} is restricted to
 the lowest order two-loop graphs in the action, leading to bubble diagram
resummation, the formalism can be extended to
include higher order graphs \cite{Verschelde:2001dz,Verschelde:2000ta,Smet:2001un}.
These approaches can be taken over to a formulation of nonequilibrium
equations of motion.  Here we restrict ourselves to including
two-loop graphs of leading order in $\lambda$ and $1/N$ only;
within the CJT formalism this is denoted as Hartree approximation
\cite{Cornwall:1974vz}.

Having at our disposal a formalism for renormalized finite-$N$ nonequilibrium dynamics
we can study the new features
introduced by the back reaction between  quantum fluctuations
in scalar quantum field theories with spontaneous symmetry breaking
for {\em finite $N$}.
A case of particular interest is  $N=4$, the $O(4)$ sigma model
is widely used as an effective theory for low energy meson interactions.
Its nonperturbative aspects may be an essential element for
understanding the phenomena observed at  RHIC \cite{Muller:1999wa}.
 Such aspects have been discussed previously in
\cite{Gavin:1993px,Mrowczynski:1995at,Kaiser:1998hf,Hiro-Oka:1999xk,GomezNicola:2001js}.
Larger values of $N$ may be realized in grand unified theories, whose nonequilibrium
evolution may be of importance in inflationary cosmology \cite{Linde:2000kn,
Dolgov:1982th, Abbott:1982hn,Kofman:1994rk,Boyanovsky:1994xf,
Shtanov:1994ce,Boyanovsky:1996rw,
Ramsey:1997sa}; we have
included simulations for a suggestive value $N=10$.
It should be stressed that the present approximation just constitutes a
$1/N$ {\em correction}, and the application to low values of $N$
should be taken, therefore, with caution. Indeed, higher order corrections
obtained when including the sunset diagram have been found to
be of importance \cite{Smet:2001un}  in equilibrium
quantum field theory. In  nonequilibrium
quantum field theory the r\^ole of such higher corrections is being
discussed at present \cite{Mihaila:2000ib,Mihaila:2001sr,Blagoev:2001ze,
Aarts:2000mg,Berges:2000ur,Berges:2001fi,Aarts:2001qa}.

The out-of-equilibrium configuration
that has mainly been studied is characterized by an initial
state in which one of the components has a
spatially homogenous classical expectation value
$\phi(t)$. For brevity, and referring to the $O(4)$ case used
as a model of low energy pion interactions, we call this component
sigma ($\sigma, a=1$), the remaining $N-1$ components pions ($\pi, a=2\dots N$).
In the initial state $\sigma$ has a classical value $\phi(0)$, the quantum vacuum
is characterized by a Bogoliubov transformation of the Fock space
vacuum state, a Bogoliubov transformation characterized by initial
masses $\calm_1(0)$ and $\calm_2(0)$. These masses are determined self-consistently,
as are their values at finite times.

The evolution of the system is governed by the classical equation
of motion for the field $\phi(t)$ and by the mode equations
for the quantum fields $\eta_a(\bfx,t)$. The expectation
values $\langle \eta_a(\bfx,t)\eta_a(\bfx,t)\rangle, a=1\dots N$, appear in both
equations of motion, this constitutes the quantum back reaction.
In the large-$N$ limit one omits the quantum fluctuations
of the sigma mode $a=1$ and just considers the Goldstone modes.
Here we are able to study the behavior of the $\sigma$ fluctuations as well,
we will find that this is a rather important aspect in the critical region.

The plan of this paper is as follows: in section \ref{formul} we introduce the
model and the potential obtained by bubble diagram resummation
in unrenormalized form. Renormalization is discussed  in section \ref{renorm}.
We present our numerical results in section \ref{resul} and discuss their implication
for the phase structure of the model. Conclusions are presented in
section \ref{conclus}.

%%%%%%%%%%%%%%%%%%%%%%%%%%%%%%%%%%%%%%%%%%%%%%%%%%%%%%%%%%%%%%%%%%%%%%%%%%%%%%%%

\section{\label{formul}Formulation of the model}
\setcounter{equation}{0}
We consider the $O(N)$ vector model with spontaneous
symmetry breaking as defined by the Lagrange density
\be
  \call=\frac{1}{2} \del_\mu \vec{\Phi}\,
  \del^\mu \vec{\Phi} -
  \frac{\lambda}{4} \left( \vec{\Phi}^2 - v^2 \right)^2
\pkt\ee
We consider a quantum system out of equilibrium that is characterized by
a spatially homogenous background field.
The fields are separated as
\be
\Phi_a= \phi_a(t) + \eta_a(\bfx,t)
\ee
into a classical part
$\phi_a = \langle \Phi_a \rangle$ and a fluctuation part
$\eta_a$ with $\langle \eta_a \rangle=0$.
Furthermore, in view of  spatial translation invariance
it is convenient to decompose the quantum fluctuations
via
\be
\eta_i(\bfx,t)=\intki \left[ a_\bfk f_i(k,t) e^{i\bfk\bfx}
+a^\dagger_\bfk f^*_i(k,t) e^{-i\bfk\bfx}\right]\kma
\ee
where $\omega_{0i}=\sqrt{k^2+m_{0i}^2}$. $m_{0i}$ will be defined below.
The subscript $i=1$ denotes the sigma mode ($a=1$),
$i=2$ denotes the pion modes ($a=2\dots N$).

In formulating the equations of motion and the renormalization
we follow the presentation of Nemoto et al. \cite{Nemoto:2000qf}
whose generalization to the nonequilibrium system is straightforward.

We introduce the inverse propagator in the classical background
field in an $O(N)$ symmetric form
\be \label{invprop}
\calg_{ab}^{-1}=
\left[\Box + \calm_2^2\right]\delta_{ab}+
\frac{\phi_a\phi_b}{\vec\phi^2}\left[\calm_1^2(t)-\calm_2^2(t)\right]
\pkt\ee
Here $\calm_{1,2}$ are trial masses that will be determined self-consistently.
In contrast to equilibrium quantum field theory these masses,
as well as the classical field, are allowed to
depend on time. This is not the most general parametrization
of an inverse propagator; it is sufficiently general if the CJT
formalism is restricted to the Hartree-Fock approximation,
but not beyond it
(see, e.g. \cite{Berges:2001fi}). If the inverse propagator
has this restricted form, the propagator itself can be
written in factorized form in terms of the mode
functions $f_i(k,t)$.  In our application the classical field has just one
nonvanishing component $\phi_1$; then the inverse propagator has only diagonal
elements and these read
\be \label{CJTgreen}
\calg_{ii}(x,x')=\intki \exp\left(i\bfk\cdot(\bfx-\bfx')\right)
f_i(k,t_>)f_i^*(k,t_<)
\ee
(no summation over $i$). If the classical field has more than one nonvanishing
component, the Green function can be likewise expressed in terms of mode
functions of a coupled system \cite{Cormier:2001iw,Ba_Mi_inprep}. The occurrence of the mode functions
allows for an interpretation in terms of Fock space states which has been
used often to define particle numbers that refer explicitly to such
a basis. If higher order corrections are included such naive interpretations
have to be reconsidered.

Superficially $\calm_1$ can be associated with the sigma mass, and
$\calm_2$ with the pion mass. The actual meaning of these mass type
parameters is more subtle, as discussed by Nemoto et al.. We will come
back to this point later on.

Using the ansatz \eqn{invprop} for the propagator Nemoto et al.
derive the CJT effective action, which for a space and time independent
configuration is characterized by the effective potential
\bea \nonumber
V(\phi,\calm_1,\calm_2)&=&
\frac{1}{2} \calm_1^2\phi^2-\frac{\lambda}{2} \phi^4 +
\frac{1}{2\lambda (N+2)}m^2\left\{\calm_1^2 +(N-1)\calm_2^2\right\}
\\ \nonumber
&&-\frac{1}{8\lambda(N+2)}\left[
(N+1)\calm_1^4+3(N-1)\calm_2^4-2(N-1)\calm_1^2\calm_2^2+2 N m^4\right]
\\
&&+\frac{1}{2}\int\frac{d^4 k}{(2\pi)^4}\ln(k^2+\calm_1^2)
+\frac{N-1}{2}\int\frac{d^4 k}{(2\pi)^4}\ln(k^2+\calm_2^2)
\pkt\eea
Here $\phi^ 2=\vec \phi^ 2$.
We note that our convention for the coupling constant differs from the one
in \cite{Nemoto:2000qf}, furthermore we have to
set $m^2=-\lambda v^2$.

We can easily generalize this effective potential to obtain the nonequilibrium
energy density
\bea
\cale &=&\frac{1}{2} \dot \phi^2 +
\frac{1}{2} \calm_1^2\phi^2-\frac{ \lambda}{2}² \phi^4 -
\frac{1}{2 (N+2)}v^2\left\{\calm_1^2 +(N-1)\calm_2^2\right\}
\\ \nonumber
&&-\frac{1}{8\lambda(N+2)}\left[
(N+1)\calm_1^4+3(N-1)\calm_2^4-2(N-1)\calm_1^2\calm_2^2+2 N \lambda^2 v^4\right]
\\ \nonumber
&&+\cale_{{\rm fl},1} + (N-1) \cale_{{\rm fl},2}
\eea
with
\be \label{enint}
\cale_{{\rm fl},i} = \frac{1}{2}
\intki
\left[|\dot f_i|^2 + (k^2+\calm_i^2(t))|f_i|^2\right]
\pkt
\ee
The equations of motion can easily be derived by requiring that this energy density
be conserved. This is the case if the equation of motion for the field
$\phi$ is given by
\be
\ddot \phi(t) + \left[\calm_1^2(t)-2 \lambda \phi(t)^2\right]\phi(t)=0
\kma
\ee
if the quantum modes satisfy the equations of motion
\be \label{modeqs}
\ddot f_i(k,t)+\left[k^2+\calm_i^2(t)\right] f_i(k,t)=0
\kma
\ee
and if the trial masses satisfy, for all $t$, the gap
equations
\bea  \label{gapeq1}
\calm_1^2(t)&=&\lambda\left[3 \phi^2(t)-v^2+3 \calf_1(t)+ (N-1)\calf_2(t)\right]
\\ \label{gapeq2}
\calm_2^2(t)&=&\lambda\left[\phi^2(t)-v^2+ \calf_1(t)+ (N+1)\calf_2(t)\right]
\pkt
\eea
Here $\calf_i(t)$ are the fluctuation integrals
\be \label{flint}
\calf_i(t)=\intki |f_i(k,t)|^2
\pkt
\ee
The gap equations incorporate the resummation of bubble diagrams.

 For a time-dependent problem we have to specify initial conditions.
We choose at $t=0$ a value of the classical field $\phi_0=\phi(0)$ different
from its value in the equilibrium ground state, which is given by
$v$ apart from quantum corrections.
The initial mass parameters $m_{i0}=\calm_i(0)$ are obtained by solving the gap equations
\eqn{gapeq1} and \eqn{gapeq2} at $t=0$, i.e., by finding the extremum of the effective
potential at a fixed value of $\phi=\phi_0$. So the initial
configuration is an equilibrium configuration with an externally fixed field $\phi_0$.
When the field is allowed, for $t > 0$, to become an internal dynamical field
 the nonequilibrium evolution sets in.

The equations of motion and the gap equations still contain divergent
integrals and need to be replaced by renormalized ones. These will be
derived in the next section.

Before continuing in developing the formalism we would like to come back
to the discussion of the masses. The mass parameters $\calm_i$
naively represent  the effective masses for the $\sigma$ and
$\pi$ fluctuations. In finite temperature
quantum field theory one expects massless quanta, the Goldstone modes,
if the field is in the temperature-dependent minimum of the
effective potential in the broken symmetry phase.
Likewise, in the nonequilibrium evolution of large-N systems
it has been found \cite{Cooper:1997ii,Boyanovsky:1996sq,
Boyanovsky:1998zg,Boyanovsky:1999yp,Baacke:2000fw}
that the mass of the fluctuations goes to zero in the broken symmetry phase as
the classical field approaches an equilibrium value. In some sense that
is trivial there, because the mass of the fluctuations appears in
the classical equation of motion as well, and in equilibrium we must have
$\ddot \phi=-\calm^2(t)\phi=0$.
Here the situation is more complicated and indeed we will find that
$\calm_2^2(t)$ will {\em not} go to zero at late times, even if the
classical field settles at some constant value.
Likewise, in the static case, at finite temperature,
one finds that the
``pion'' mass $\calm_2$ is not in general zero, the question
has been rised, whether Goldstone' s theorem
is violated by the approximation or otherwise.

The interpretation of the masses and the value of the
pion mass in the present context have been discussed extensively in
Refs.\cite{Nemoto:2000qf, Verschelde:2000ta}.
These authors argue that $\calm_2$ is {\em not} to be interpreted as the
pion mass, but as a variational parameter which does not necessarily
have an immediate physical interpretation. The ansatz for the inverse Green function
and in consequence the effective potential have full $O(N)$ symmetry.
The field $\phi$ appearing there is the {\em absolute value} of
the field $\phi_a$. Therefore by symmetry the pion mass must
trivially be zero
in the minimum of the effective potential, which is an $O(N-1)$
sphere.
Nemoto et al. show that trivially the appropriate second derivatives of
the effective potential lead to a vanishing pion mass. Likewise
$\calm_1$ is another parameter characterizing the Green function
and is different from  the sigma
mass which is given by
\be \label{sigmamass}
m_\sigma^2=2v^2(\frac{d\calm_1^2}{d\phi^2}-2\lambda)
\pkt\ee
This is equal to $2 \lambda v^2$ on the tree level only.

It is instructive to do the simple algebra of taking second derivatives
of an $O(N)$ symmetric function:
\be
\frac{\partial^2 f(\vec \phi^2)}
{\partial \phi_a \partial\phi_b}=2 \delta_{ab}
f'(\vec\phi^2)+4 \phi_a\phi_b f''(\vec\phi^2)
\pkt\ee
The pion mass is the second derivative perpendicular to the direction
of $\phi_a$, so it is given by the first term and vanishes if
$f'(\vec \phi^2) =0$, which defines the minimum. This corresponds
precisely to the {\em first} derivative of the potential appearing in the
classical equation of motion.
 We note that this effective mass is given by
\be \label{classmass}
\calm_{\rm cl}^2=\biggl(\calm_1^2(t)-2 \lambda \phi^2(t)\biggr)
\pkt\ee
It vanishes trivially if the field settles at late times at some
constant value.

 The masses $m_\sigma^2$ and $\calm_{\rm cl}^2$ determine the fluctuations
of the classical field $\phi$ near the minimum of the effective
potential in analogy to the tree level masses
of the sigma and pion fields.
If all higher corrections were included, one
would expect these masses to determine the exact propagator  near $k^2=0$.
In this sense the vanishing of $\calm_{\rm cl}^2$ entails a pole
of the pion propagator at $k^2=0$. However, contrary to the large-$N$
limit, $\calm_{\rm cl}^2$ is not the mass that determines the
fluctuations $f_2$ and, thereby, the ``pion'' propagator $\calg_{22}$.

%%%%%%%%%%%%%%%%%%%%%%%%%%%%%%%%%%%%%%%%%%%%%%%%%%%%%%%%%%%%%%%%%%%%%%%%%%%%%%%%%%%

\section{\label{renorm}Renormalization}
\setcounter{equation}{0}
The renormalization of the effective potential has been discussed in
\cite{Nemoto:2000qf} and  \cite{Verschelde:2001dz}. As stated in the
latter publication both formulations are equivalent, here we follow the one of
Nemoto et al., employing the auxiliary field method, in which the counterterms
in the effective potential are introduced via the
the trial masses:
\be \label{counterterms}
\delta \cale =  A v^2\calm_1^2+B v^2  \calm_2^2+
\frac{1}{2}C \calm_1^4 + \frac{1}{2}(N-1) D \calm_2^4
\pkt
\ee
Divergent and finite parts of the fluctuation integrals $\calf_i$ and of the
fluctuation energy densities have been analyzed
in \cite{Baacke:1997se,Baacke:1998zy} using a perturbation
expansion in the ``potentials''
\be
\calv_i(t)=\calm_i^2(t)-\calm_i^2(0)
\pkt\ee
 As the definition of the Green functions \eqn{CJTgreen}, the
expressions for the energy densities \eqn{enint} and for the fluctuation
integrals \eqn{flint}, as well as the equations
\eqn{modeqs}  satisfied
by the mode functions are entirely analogous to those in \cite{Baacke:1998zy},
the expansions derived there can be applied here in the same way.
 In the power series expansion of the energy
densities  with respect to $\calv_i$ the zeroth, first and second
order terms are UV divergent, while in the fluctuation
integrals it is the zeroth and
first order terms. In dimensional regularization
the powers of $\calv_i$ can be arranged with
powers of $m_{i0}^ 2$ into powers of $\calm_i^ 2$.
The  expansion then reads
\bea \label{epert}
\cale_{{\rm fl},i}&=&\frac{1}{64 \pi^2}
\left[ \calm_i^4 (-L+\ln\frac{m_{i0}^2}{\mu^2})-2 \calm_i^2
m_{0i}^2+\frac{1}{2}m_{i0}^4\right] +
\cale_{{\rm fl},i}^{ \rm fin}
\\\label{flpert}
\calf_i&=&\frac{1}{16 \pi^2}
\left[ \calm_i^2\left(-L+\ln\frac{m_{i0}^2}{\mu^2}\right)-m_{0i}^2\right]
+\calf_{i}^{ \rm fin}
\pkt
\eea
Here
\be
L= \frac{2}{\epsilon}-\gamma+\ln 4\pi
\kma\ee
$\mu$ is the renormalization scale, and $m_{i0}$ are the mass parameters appearing
in the fluctuation integrals.
In order to avoid bad initial singularities
\cite{Cooper:1987pt,Maslov:1998bf,Baacke:1998zz} they have to be chosen as
$m_{i0}=\calm_i(0)$, the ``initial masses''.
Finally the finite fluctuation integrals are defined by subtracting the UV divergent
parts under the momentum integral via
\bea
\cale_{{\rm fl},i}^{ \rm fin}&=&
\frac{1}{2}\intki \left[|\dot f_i|^2 + (k^2+\calm_i^2)|f_i|^2\right.
\\ \nonumber && \hspace{10mm}
\left.-2 \omega_{i0}^2-\calv_i+
\frac{\calv_i^2}{4\omega_{i0}^2}\right]
\\
\calf_{i}^{ \rm fin}&=&
\frac{1}{2}\intki \left[|f_i|^2 -1+\frac{\calv_i}{2\omega_{i0}^2}\right]
\pkt
\eea
 The subtractions used here are analytically equivalent to
a numerically more sophisticated procedure used in \cite{Baacke:1997se}.
The subtracted integrals  are UV finite.

 From Eqs. \eqn{epert} and \eqn{flpert}  it is evident that
the divergent parts $\propto L$ are independent of the initial masses $m_{i0}^2$
and thereby of the initial conditions.
Exactly as in equilibrium theory \cite{Nemoto:2000qf} the divergent parts
can be removed by choosing the counter terms \eqn{counterterms} as  $A=B=0$ and
as
\be
C=D=\frac{1}{64 \pi^2} L
\pkt
\ee
With this  choice
 the energy density as well as the gap equations
become finite. The equations of motion are not affected.
The energy density is obtained simply by replacing the fluctuation
energies by the right hand sides of \eqn{epert} omitting the term
proportional to $L$. Likewise the renormalized gap
equations are obtained by replacing the fluctuation integrals
by the right hand sides of \eqn{flpert} without the term proportional
to $L$. In our numerical computations the renormalization
scale $\mu$ has been taken equal to the sigma mass. As long as the ratio
of $ \mu$ and the relevant masses is far from the Landau pole, i.e.,
$m^ 2/\mu^ 2 \ll \exp 16 \pi^ 2/\lambda$ the dependence on the
renormalization scale is weak. This is the condition under which
the $\Phi^4$ theory can  be treated as a low energy effective theory.
 For our numerical simulations with
$\lambda=1$ this condition is fulfilled.

The gap equations for the masses have to be solved once, at $t=0$.
By subtracting these equations at $t=0$ from the general gap equations
one obtains the renormalized gap equations for the potentials $\calv_i$.
These read explicitly
\bea
\calv_1&=&\lambda \left[3(\phi^2-\phi_0^2)+3 \calf_1^{\rm fin}+
(N-1)\calf_2^{\rm fin}\right.
\\ \nonumber &&\left. +\frac{3}{16 \pi^2}\ln\frac{m_{10}^2}{\mu^2}
\calv_1(t)+\frac{N-1}{16 \pi^2}\ln\frac{m_{20}^2}{\mu^2}
\calv_2(t)\right]
\\
\calv_2&=&\lambda \left[(\phi^2-\phi_0^2)+ \calf_1^{\rm fin}+
(N+1)\calf_2^{\rm fin}\right.
\\ \nonumber &&\left. +\frac{1}{16 \pi^2}\ln\frac{m_{10}^2}{\mu^2}
\calv_1(t)+\frac{N+1}{16 \pi^2}\ln\frac{m_{20}^2}{\mu^2}
\calv_2(t)\right]
\pkt
\eea
These linear equations can be solved easily for $\calv_i(t), i= 1,2$
using a time-independent matrix. This matrix is analogous to the
factor $\calc=(1+(\lambda/16 \pi^2)\ln(m^2/m_0^2))^{-1}$ appearing
in the large-$N$ case \cite{Baacke:1998zy}.

The numerical implementation has been described in several
previous publications (see e.g. \cite{Baacke:1997se}),
so we do not repeat this here. The accuracy of the computations is monitored
by veryfying the energy conservation, which holds with a typical precision of five
significant digits.

We finally mention the problem of initial singularities that appears in the context of
renormalization \cite{Cooper:1987pt,Maslov:1998bf,Baacke:1998zz}. These can be avoided
by modifying the initial quantum ensemble via a suitable Bogoliubov transformation. This can be
done in the present model as well. For the values of the couplings and initial
parameters considered here the initial singularities are numerically unimportant and have
been disregarded, therefore.

%%%%%%%%%%%%%%%%%%%%%%%%%%%%%%%%%%%%%%%%%%%%%%%%%%%%%%%%%%%%%%%%%%%%%%%%%%%%%%%%%%%%%%

\section{\label{resul}Results and discussion}
\setcounter{equation}{0}

\subsection{\label{numsim}Numerical simulations}
We have performed numerical simulations for the cases
$N=1$, $N=4$ and $N=10$. The coupling constant was taken to be
$\lambda=1$ and we have varied the initial value for the
field $\phi_0=\phi(0)$. We have considered only values $\phi_0 > v$,
as for smaller values the initial mass $m_{20}=\calm_2(0)$ becomes
imaginary. The region $\phi < v$ is only explored dynamically.
We display the time evolution of the classical amplitude $\phi(t)$ for
the parameters $N=4, \lambda=1$ in Figs. 1a-c, 
for initial
amplitudes $\phi_0=1.3v < \sqrt{2} v$, $\phi_0=1.445 v \simeq \sqrt{2} v$
and $\phi_0=1.6 v > \sqrt{2} v$.

\setcounter{fig2}{1}
\begin{figure}[htbp]
  \begin{center}
    \sffamily
    \label{fig:1a}
    \includegraphics[scale=0.4,angle=270]{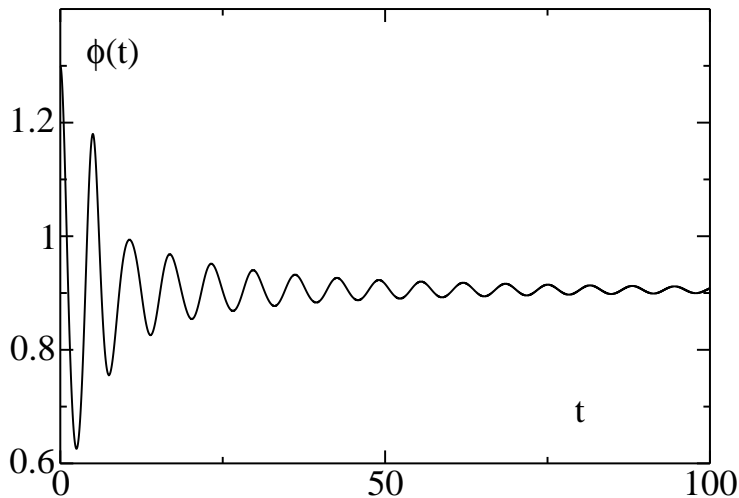}
    \caption{Time evolution of the classical field $\phi$
for $\phi_0 < \sqrt{2}v$ (broken symmetry phase). Field amplitude and time
are in units of $v$ and $v^{-1}$, respectively.
Parameters are $N=4$, $\lambda=1$, and $\phi_0=1.3v$.}
  \end{center}
\end{figure}
\addtocounter{fig2}{1}
\addtocounter{figure}{-1}

For the tree level potential the value $\phi=\sqrt{2} v$ is the value for which
the energy is equal to the height of the barrier ($N=1$) or the top
of the Mexican hat ($N>1$). We will discuss the physics associate with
the three ranges of parameters below.

\setcounter{fig2}{2}
\begin{figure}[htbp]
  \begin{center}
    \sffamily
    \label{fig:1b}
    \includegraphics[scale=0.4,angle=270]{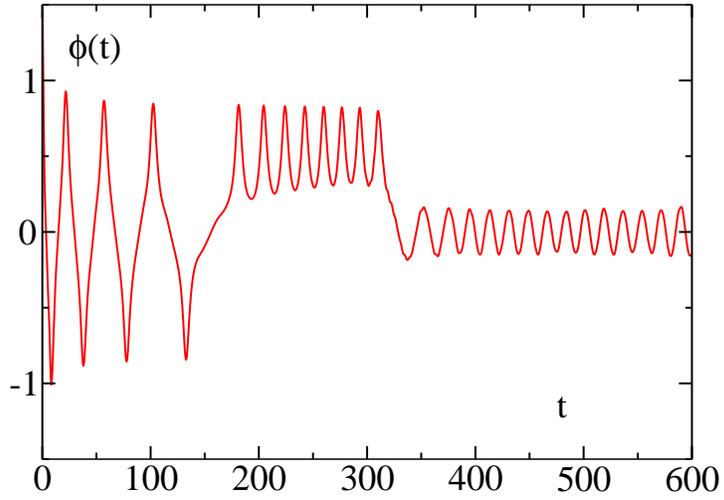}
    \caption{ Time evolution of the classical field $\phi$
for $\phi_0 \simeq \sqrt{2}v$ (critical region). Parameters as above but
$\phi_0=1.445v$.}
  \end{center}
\end{figure}
\addtocounter{fig2}{1}
\addtocounter{figure}{-1}

\setcounter{fig2}{3}
\begin{figure}[htbp]
  \begin{center}
    \sffamily
    \label{fig:1c}
    \includegraphics[scale=0.4,angle=270]{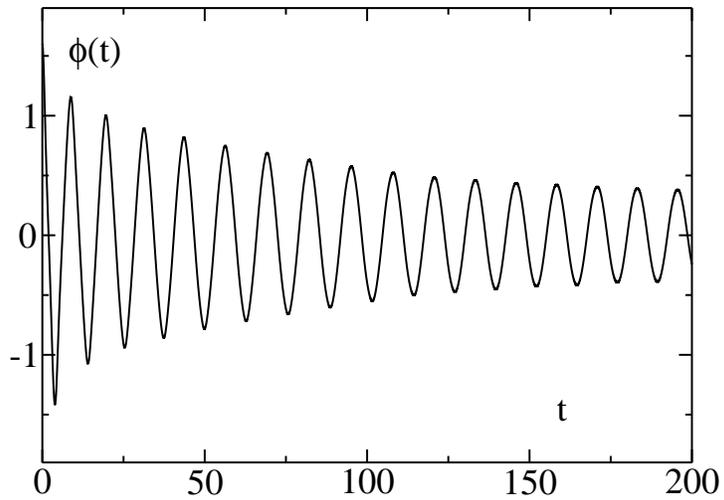}
    \caption{ Time evolution of the classical field $\phi$
for $\phi_0 > \sqrt{2}v$ (symmetric phase). Parameters as above but
$\phi_0=1.6v$.}
  \end{center}
\end{figure}
\setcounter{fig2}{1}

In the nonequilibrium evolution of models with spontaneous
symmetry breaking both squared masses $\calm_{1,2}^2$ can in general take
negative values. In large-$N$ dynamics it is well known
\cite{Boyanovsky:1998zg,Boyanovsky:1999yp,Baacke:2000fw}
that in this situation the fluctations
increase exponentially and drive the  squared masses back to
positive values. This stabilizes the system dynamically and
prevents an unphysical behavior in which an exponentially increasing
amount of quantum energy, taken from the vacuum, is converted into classical one.
Our first essential observation is that for all parameter sets
and initial values this stabilization takes place for finite
$N$ as well. We display a typical time evolution of the classical field
and of the squared masses $\calm_{1,2}^2$ in Fig. 2a and Fig. 2b,
based on the parameters $N=4, \lambda=1$, and $\phi_0=1.4 v$.
In Fig. 2a both squared masses are seen to become negative at early times
and to reach  positive values at late times. In Fig. 2b one sees
the evolution of the energy of the fluctuations which is very strong in the
time range where the squared masses are negative.

\setcounter{fig2}{1}
\begin{figure}[htbp]
  \begin{center}
    \sffamily
    \label{fig:2a}
    \includegraphics[scale=0.4,angle=270]{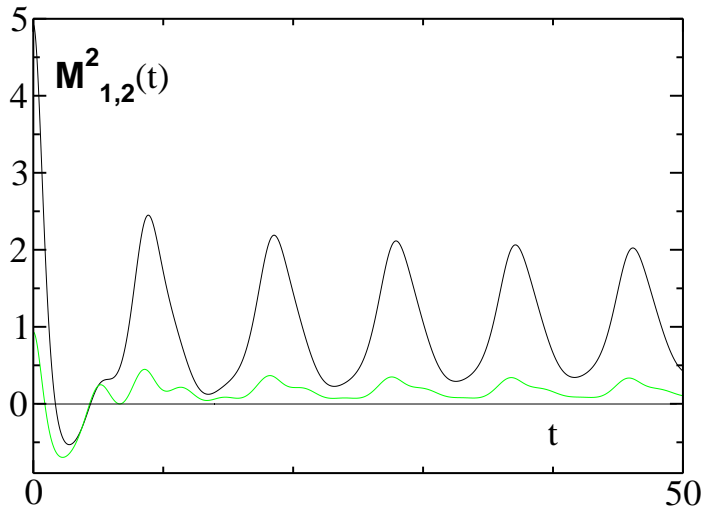}
    \caption{ Time evolution of the masses $\calm_{1,2}^ 2$.
Parameters are $N=4$,  $\lambda=1$, and $\phi_0=1.4v$. The upper curve
is $\calm_1^2$.}
  \end{center}
\end{figure}
\addtocounter{figure}{-1}
\addtocounter{fig2}{1}

\begin{figure}[htbp]
  \begin{center}
    \sffamily
    \label{fig:2b}
    \includegraphics[scale=0.4,angle=270]{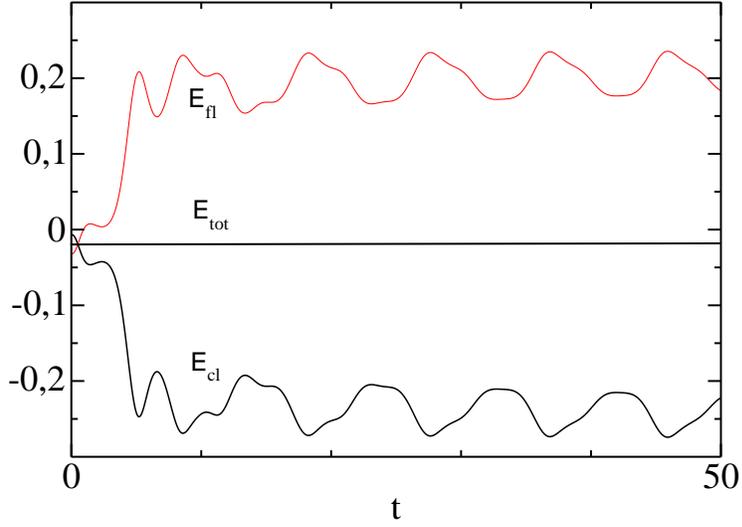}
    \caption{ Time evolution of the energy densities.
Parameters as in  Fig. 2a. Displayed are the classical energy density,
the fluctuation energy density including finite renormalization terms and
the total energy density.}
  \end{center}
\end{figure}
\addtocounter{figure}{-1}
\addtocounter{fig2}{1}

{The classical minimum of the energy
is at $E_0=-\lambda v^4/4$, i.e., $E_0 = -.25$ for this parameter set.
 So essentially all the energy is transferred to fluctuations
within a few oscillations of the condensate field $\phi$.
In Fig. 2c we display the two fluctuation integrals as functions of time
for $\phi_0=2v$, again for $N=4$ and $\lambda=1$, i.e. in the symmetric phase.
The pion fluctuations are seen to develop rather quickly, while the
sigma fluctuations only develop at later times.}

\begin{figure}[htbp]
  \begin{center}
    \sffamily
    \label{fig:2c}
    \includegraphics[scale=0.4,angle=270]{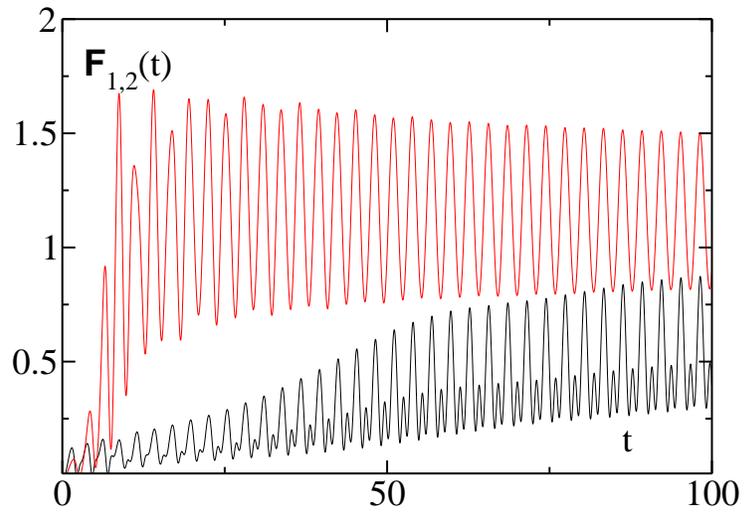}
    \caption{Time evolution of the fluctuations integrals.
Parameters as in  Fig. 2a, however with $\phi_0=2v$. upper curve:
pion fluctuations, lower curve: sigma fluctuations.}
  \end{center}
\end{figure}
\setcounter{fig2}{1}

The main aspect we will consider here is the phase structure of the model.
In thermal quantum field theory one expects a
phase with spontaneously broken symmetry at low temperature, and
symmetry restoration at high temperature. Here we have a microcanonical
description, so instead of temperature we specify the energy which in turn
is determined by the initial value $\phi_0$. As already mentioned the system
will not thermalize in the approximation used here, but it is characterized
at late times by  limiting values,  or time averages, attained by the
various masses, and by the classical field $\phi$. The latter can be compared with
the temperature-dependent vacuum expectation value $v(T)$, the former ones
are associated to correlation lengths.

\setcounter{fig2}{1}
\begin{figure}[htbp]
  \begin{center}
    \sffamily
    \label{fig:3a}
    \includegraphics[scale=0.4,angle=270]{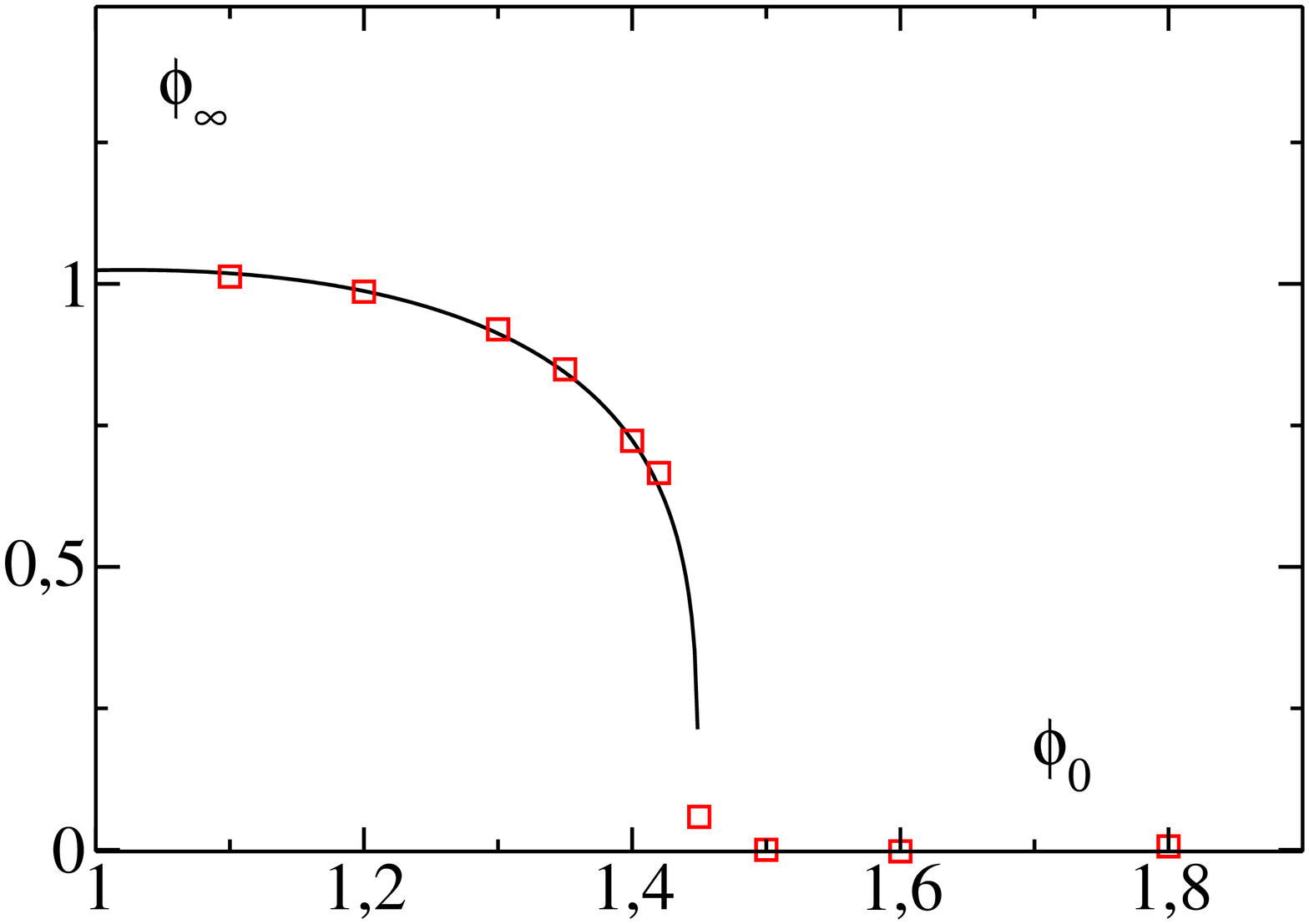}
    \caption{The late-time amplitude as a function of the
initial amplitude. Parameters are $N=1$ and $\lambda=1$.}
  \end{center}
\end{figure}
\addtocounter{figure}{-1}
\addtocounter{fig2}{1}

\begin{figure}[htbp]
  \begin{center}
    \sffamily
    \label{fig:3b}
    \includegraphics[scale=0.4,angle=270]{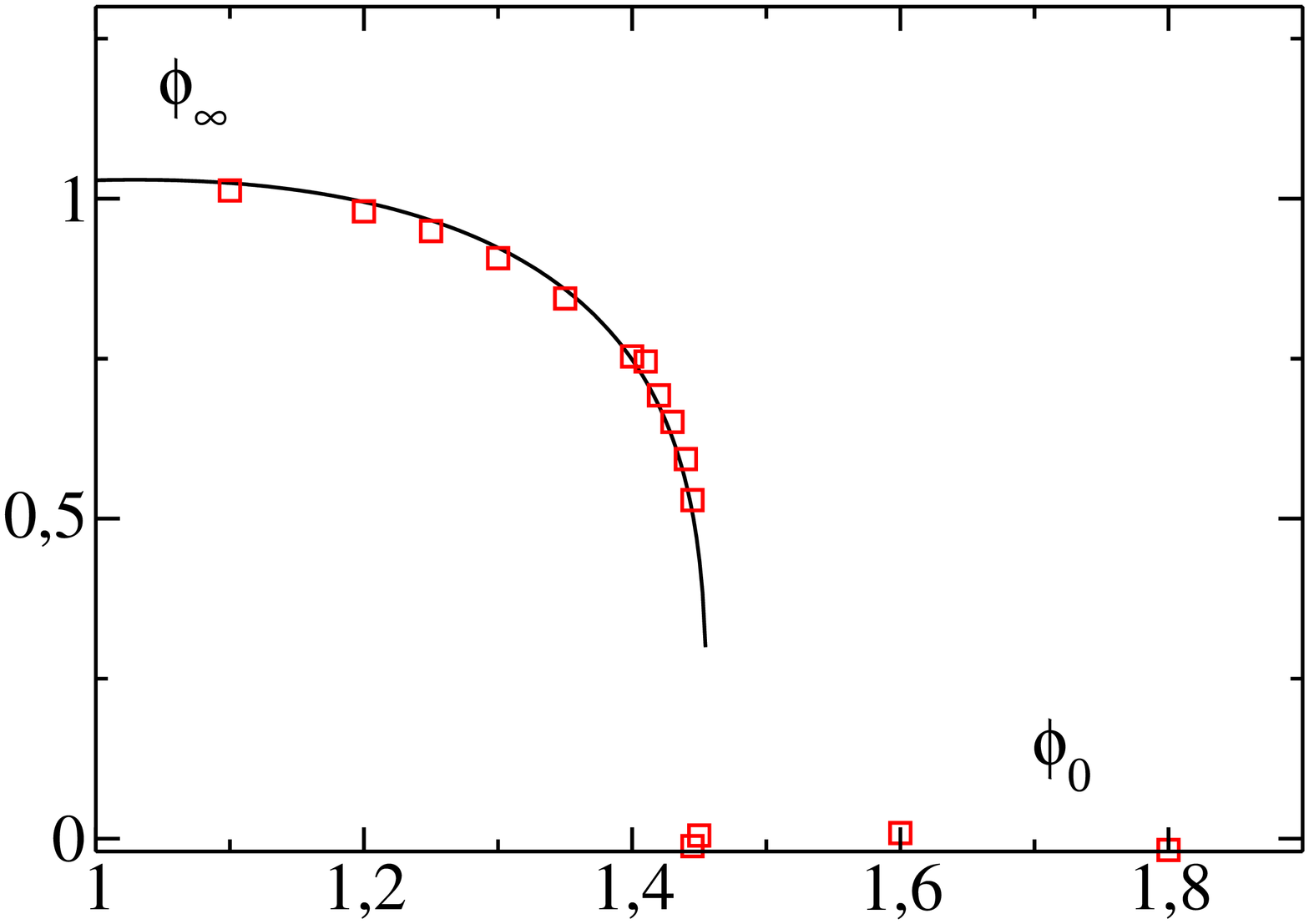}
    \caption{ Same as Fig. 3a, for $N=4$ and $\lambda=1$.}
  \end{center}
\end{figure}
\addtocounter{figure}{-1}
\addtocounter{fig2}{1}

\begin{figure}[htbp]
  \begin{center}
    \sffamily
    \label{fig:3c}
    \includegraphics[scale=0.4,angle=270]{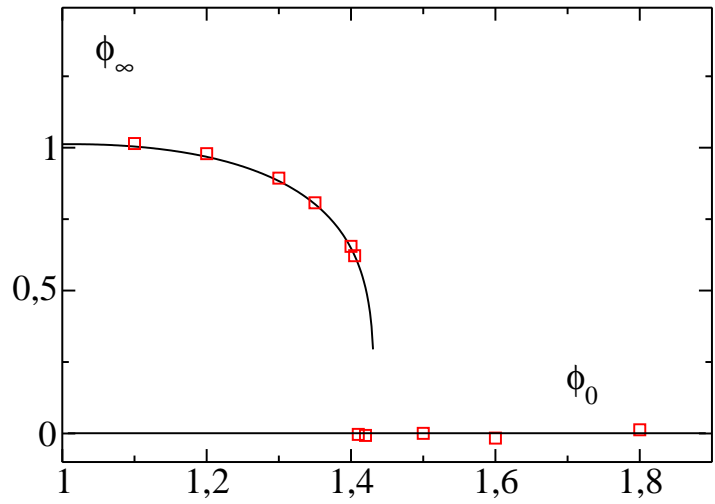}
    \caption{ Same as Fig. 3a, for $N=10$ and $\lambda=1$}
  \end{center}
\end{figure}

Classically the spontaneously broken phase is expected to be situated in
the interval $v < \phi_0 < \sqrt{2} v$. For larger values of $\phi_0$ one
should be in the phase of restored symmetry.
Since the effective potential receives quantum corrections
the boundaries of these intervals will be slightly shifted, for brevity we
continue to use the classical values in the following discussion.

\subsection{\label{classfield}Late time behavior of the classical field}

We first consider $\phi_\infty(\phi_0)$, the value of $\phi$ averaged at late times, which
can be considered as an order parameter, replacing the vacuum expectation value
$v(T)$ of finite temperature field theory.
In the analysis of the broken symmetry phase in the large-$N$ limit
it was observed \cite{Boyanovsky:1998zg, Boyanovsky:1999yp,Baacke:2000fw} that the value
of $\phi$ at late times has a specific and universal
dependence on the initial value $\phi_0$, given at zero temperature by
\be \label{universal}
\phi_\infty \simeq \left[\phi_0^2\left(2 v^2 - \phi_0^2\right)
\right]^{1/4}
\ee
 We find that this is the case for finite $N$ as well. The dependence
$\phi_\infty(\phi_0)$ is shown in Figs. 3a-c for $N=1,4$ and $10$,
respectively, the curves are rough fits of the form of Eq. \ref{universal},
taking into account
a slight shift of the point of ``phase transition'' $\phi\simeq \sqrt{2} v$.
Such a functional behavior is typical of a phase transition of second order.
In a first oder phase transition the vacuum expectation value of the true minimum
jumps at the phase transition from the broken symmetry minimum to the
symmetric one. As the functional behavior of Eq. \eqn{universal}
has an infinite slope at the critical point it is
 hard to decide from numerical data whether   $\phi_\infty$ near $\phi_0=
\sqrt{2}v$ goes to zero continously or via a discontinuity.
The results strongly indicate that the transition is discontinuous.
This is suggested as well by the time evolution of $\phi$ displayed in
Fig. 1b. There seems to be a minimum at $\phi=0$ and another
one near $\phi\simeq .5 v$, as one knows it from typical
finite temperature potentials of first order phase transitions.
However, even if the transition is first order,
it is very close to a second order one.
In thermal equilibrium a first order phase
transition is expected \cite{Nemoto:2000qf,Smet:2001un}, it becomes second
order only after including higher loop corrections.
Fig. 1b puts in evidence that the time  averaging is problematic near the
phase transition; this implies that the functional dependence the ``order parameter''
$\phi_\infty$ near $\phi_0 \simeq \sqrt{2}v$ is not determined with high precision.

 We would like to remark that we have chosen special initial conditions
which allow the classical field to move in one fixed direction only. So the system cannot
see the difference between a double-well and a Mexican hat potential.
Imposing more general initial conditions \cite{Ba_Mi_inprep} may lead to an improved understanding
of the nonequilibrium properties of the system. Such a generalization may
be useful especially near the transition point $phi_0\simeq \sqrt{2} v$.

 For $\phi  > \sqrt{2} v$ the time average of $\phi$ vanishes as $t \to \infty$, and in fact
already at an early stage of evolution. However, the amplitude of the oscillations
decreases very slowly, if at all. So, although the order parameter $\phi_\infty$
shows the behavior expected in the symmetric phase of
a thermal system, and although the system becomes essentially stationary, it shows no
resemblance to a system in thermal equilibrium with
a constant value of $\phi$. We will further analyze this phase, below.

\subsection{\label{avmasses}Late time behavior of the masses}

Another set of variables characteristic for the phase structure of the $O(N)$ model in equilibrium
are the mass scales or correlation lengths.
In the broken symmetry
phase we expect a vanishing pion mass and a nonzero value
of the  field $\phi$. In section \ref{formul} we have discussed to some extent
the problem of the Goldstone mass, which should not be identified
naively with $\calm_2$; rather, we have proposed to associate it  with
the ``classical mass'' $\calm_{\rm cl}$ defined in  Eq. \eqn{classmass}.
Both masses would coincide in the large-$N$ limit.

\renewcommand{\thefigure}{\arabic{figure}}
\begin{figure}[htbp]
  \begin{center}
    \sffamily
    \label{fig:4}
    \includegraphics[scale=0.4,angle=270]{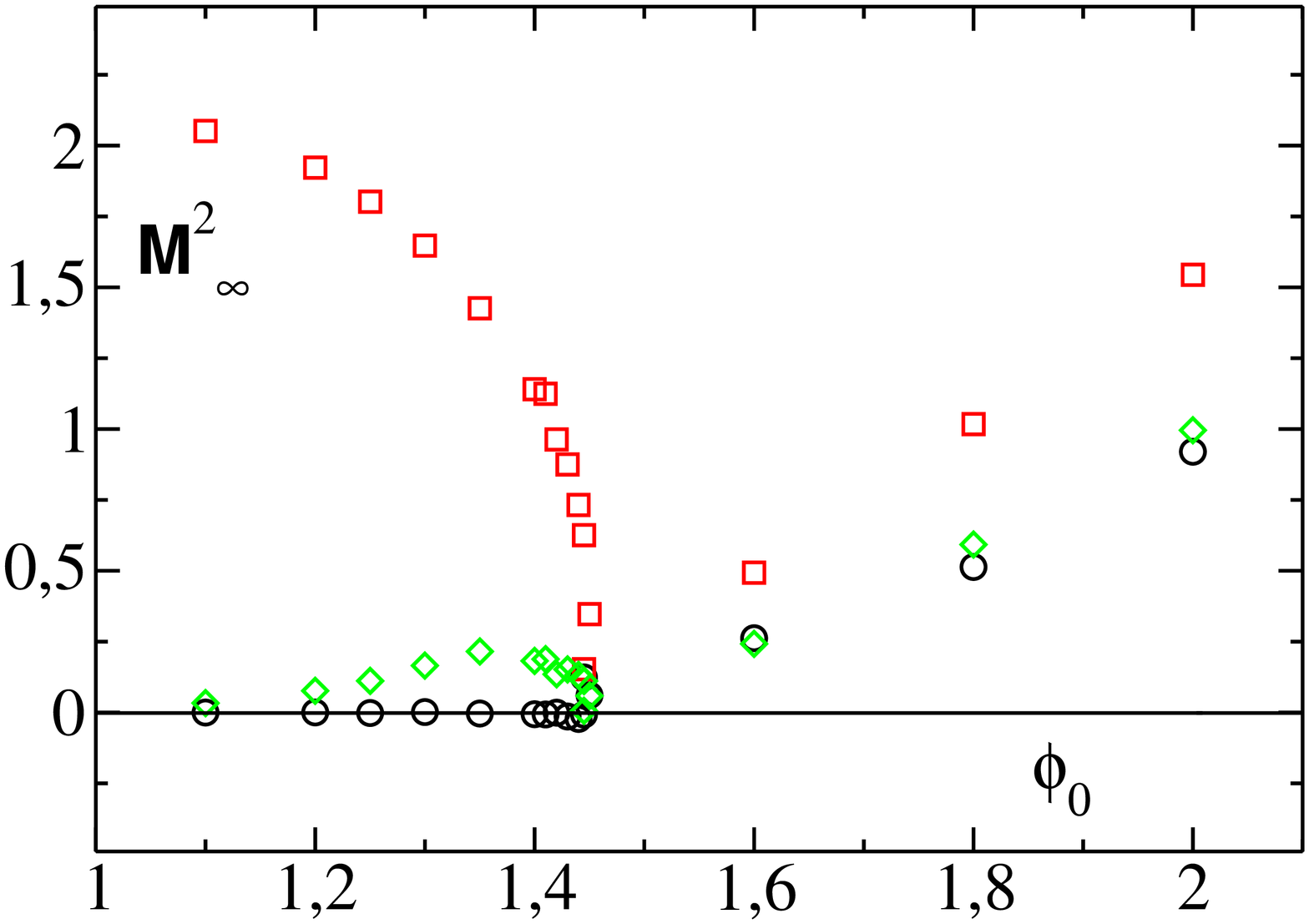}
    \caption{ The various masses averaged at late times (t=300),
 as functions of the initial amplitude.
Squares: $\calm_1^ 2$, diamonds: $\calm_2 ^2$, circles:
$\calm_{\rm cl}^ 2$. Parameters are $N=4$ and $\lambda=1$.}
  \end{center}
\end{figure}
\addtocounter{fig2}{1}

Likewise, the sigma mass is not given naively by $\calm_1$.
We plot the three squared masses $\calm_1^2$, $\calm_2^2$, and
$\calm_{\rm cl}^2$ as  functions of $\phi_0$
in Fig. 4, for $N=4, \lambda=1$. $\calm_{\rm cl}^2$ is seen to be
almost zero for  $v < \phi_0 < \sqrt{2} v$ and to increase for larger values
of $\phi_0$, as expected for the Goldstone mass. The mass
$\calm_2^2$ is small for $v < \phi_0 < \sqrt{2} v$ but definitely
different from zero, and increases likewise for $v > \sqrt{2} v$.
The ``sigma mass'' $\calm_1$ is different from zero everywhere, except
near  $\phi_0 = \sqrt{2} v$.

The formalism presented here contains of course the limit
$N \to \infty$. This limit is obtained by letting $\lambda
\propto 1/N$ and $\phi, \phi_0, v \propto \sqrt{N}$. In this limit
the quantum fluctuations
$\eta_1$ become irrelevant, and so does $\calm_1$.
The ``pion'' mass $\calm_2$ and
the ``classical mass'' $\calm_{\rm cl}$ become
identical to the mass $\calm$ of Refs. \cite{Boyanovsky:1996sq,Boyanovsky:1998zg,
Boyanovsky:1999yp,Baacke:2000fw} or $\chi(t)$ of Ref. \cite{Cooper:1997ii}.
In the broken symmetry phase the instability pushes this mass to zero.
Here, for finite $N$,  we find $\calm_2^2$
to remain positive though small in the broken symmetry phase, while
 $\calm_{\rm cl}^2$ again vanishes. It should be mentioned
that the vanishing of the classical mass $\calm_{\rm cl}^2$  in the broken symmetry phase
is due to the fact
 that the field  $\phi(t)$ settles at a  constant  nonzero value.
For the fluctuation mass $\calm_2^2$ the dynamics of backreaction only
forces it to be {\em nonnegative}. In the large-$N$ limit the identity of
the two masses $\calm^2=\calm_2^2=\calm_{\rm cl}^2$ entails their
vanishing at late times, as expected from the Nambu-Goldstone theorem.
The situation found here at large times is analogous to the one found in thermal
equilibrium \cite{Nemoto:2000qf, Verschelde:2000ta} (see section \ref{formul}).

For a phase transition of second order the $\sigma$ mass,  all
mass scales  are  expected to
vanish at the ``critical point'', i.e., for $\phi_0\simeq\sqrt{2}v $.
We see from Fig. 4 that this is almost the case.
For a first order phase transition the curvature of the
potential in the broken symmetry minimum decreases when approaching the phase transition,
but remains positive up to and beyond the phase transition. Above the phase transition the
curvature in the symmetric minimum increases again.
The fact that $\calm_1^ 2$ does not really reach zero may of course be
a deficiency of our numerics.
We  point out, however,  that  critical behavior implies large length
and also time scales, so that in this neighborhood the
time averaging is precarious and the late time values are
not very precise. It is not clear whether the field
in Fig. 1b will jump back and
forth between the various minima at later times again
and, if it does, whether this is
not a consequence of accumulated tiny numerical errors.
We again conclude that, if the phase transition is of first order, it is
``weakly first order''. While we have mostly presented results for $N=4$, we find similar
results for $N=1$ and $N=10$. For $N=1$ the mass $\calm_2$ is of course meaningless.

%In the symmetric phase all masses increase with $\phi_0$ as they do with temperature in
%finite temperature quantum field theory. The fact that they are not equal does {\em}
%imply any symmetry breaking. The ansatz for the Green function has full
%$O(N)$ symmetry, and the ``masses'' are essentially variational parameters.
%The fact that all masses increase above the phase transition is again analogous
%to finite temperature quantum field theory. Indeed the dependence on $\phi_0$
%is quadratic as it would be on temperature. In the large-$N$ limit
%this quadratic dependence is the contents of a sum rule
%\cite{Boyanovsky:1998zg,Boyanovsky:1999yp,Baacke:2000fw}.

 As we have mentioned above, in the symmetric phase the time average
of the field $\phi$ becomes zero at late times, while the field itself
continues to oscillate with essentially constant amplitude, implying
that the time average $\langle \phi^ 2\rangle$ of the squared field remains different
from zero.
This implies that
the masses $\calm_1^ 2$ and $\calm_2^ 2$ necessarily have different time averages;
indeed the classical field contributes with $3\lambda \langle \phi^ 2\rangle$
to $\langle \calm_1^ 2\rangle$ but only with $\lambda \langle \phi^ 2\rangle$
to $\langle \calm_2^ 2\rangle$. Actually even the quantum fluctuations
$\calf_1$ and $\calf_2$
have different time averages. This is contrary to what one expects in a
 symmetric equilibrium phase. The problem is inherent in the initial
conditions which necessarily require a high excitation and the choice
of direction in
$O(N)$ space. It is appropriate at this point to compare with a $\Phi^4$ model
{\em without} spontaneous symmetry breaking, i.e., with $m^2=-\lambda v^2> 0$.
There again the classical field oscillates around $ \phi=0$ with slowly decreasing amplitude.

\renewcommand{\thefigure}{\arabic{figure}}
\begin{figure}[htbp]
  \begin{center}
    \sffamily
    \label{fig:5}
    \includegraphics[scale=0.4,angle=270]{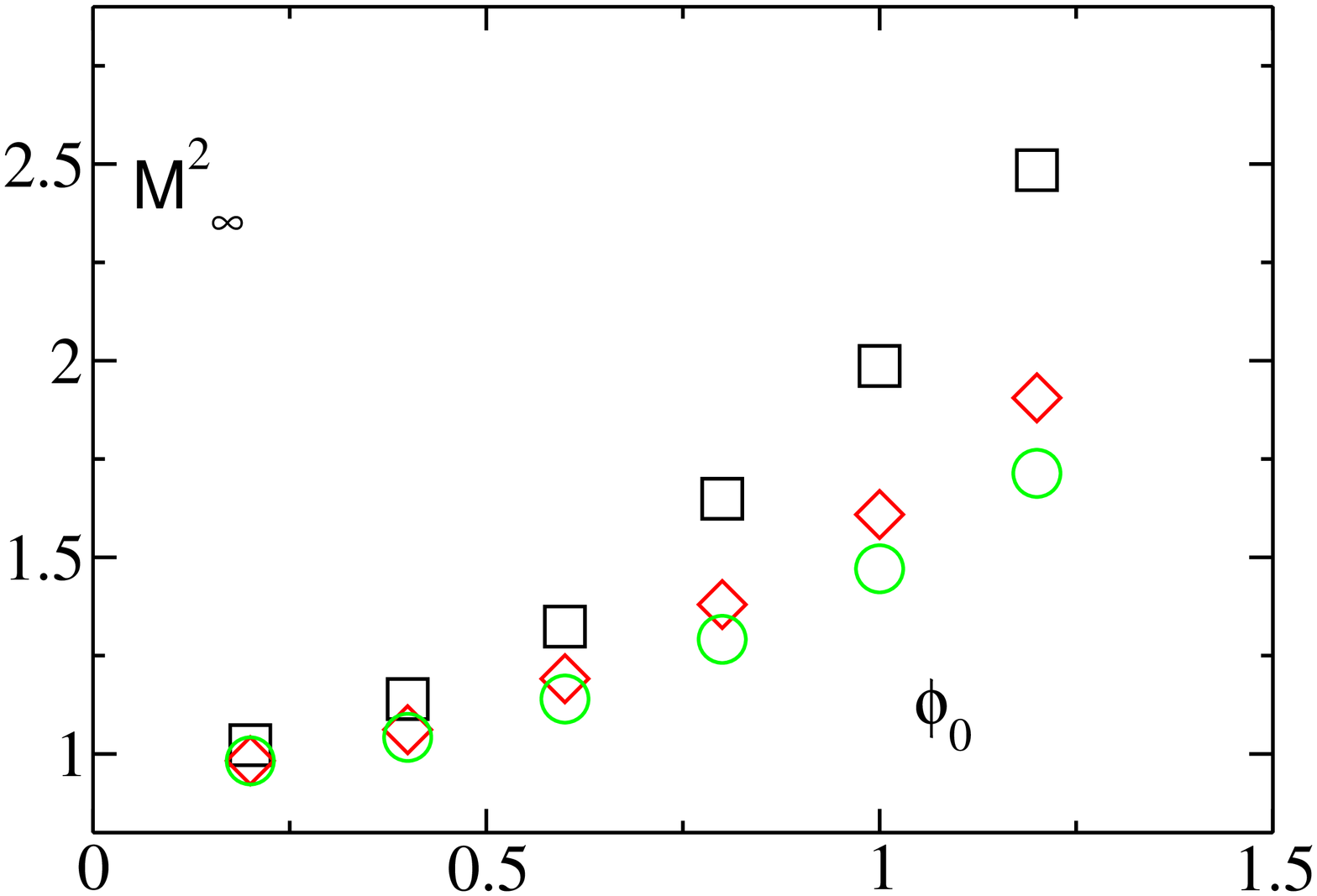}
    \caption{ The various masses averaged at late times (t=300),
as functions of the initial amplitude for a model without symmetry breaking,
$N=4$, $\lambda=1$, $v^2= -m^2/\lambda=-1$. Symbols as in Fig. 4.}
  \end{center}
\end{figure}
\addtocounter{fig2}{1}

Again the time average of mass $\calm_1^ 2$ remains different from the one
of $\calm_2^2$ pertaining to the fields $\eta_a$ with $a=2\dots N$. We show in Fig. 5
the dependence of the late time averaged masses for this manifestly symmetric theory.
The close resemblance with the behavior in the high energy phase of the model
with spontaneous symmetry breaking is obvious. The difference between the masses
is not related to spontaneous symmetry breaking but to the fact that the system
is prepared in a non-symmetric state.

While the behavior of masses $\calm_2^2$ and $\calm_{\rm cl}^2$ found here
essentially reproduces the one of $\calm^2$ in the large-$N$ limit
we have seen here, that the sigma mass contains valuable information about
the nature of the phase transition, not available in the large-$N$ limit,
and that the sigma  fluctuations are relevant for finite $N$.

\subsection{\label{correls}Correlations}

Correlations of mode functions have been discussed in various publications
\cite{Cooper:1997ii,Kaiser:1998hf,Boyanovsky:1999yp,Hiro-Oka:1999xk}. If one thinks of
the $N=4$ model as a model for pion
production, possibly displaying structures like disordered chiral condensates,
it is the correlations between the pion fluctuations ($i=2$ or $a=2\dots N$) which
are relevant. In the large-$N$ limit these are the only available correlations.
It was found in a large-$N$ computation including the
full backreaction \cite{Boyanovsky:1999yp} that the correlation length
grows with time in the broken symmetry phase. Kaiser \cite{Kaiser:1998hf},
analyzing the correlations
in the initial phase, before back reaction sets in, suggested the occurrence of
large correlation lengths. Hiro-Oka and Minakata \cite{Hiro-Oka:1999xk} performing a computation
with backreaction, however unrenormalized and with a different Hartree
factorization, suggested that the correlation length remains small.

The pion correlations are obtained \cite{Boyanovsky:1999yp} as the Fourier-Bessel transform
\bea \nonumber
 C(r,t)&=& \langle\eta_2(\bfx,t)\eta_2({\bf 0},t)\rangle
=\int\!\frac{{d^3}k }{(2\pi)^32\omega_{02}}e^{i\bfk\cdot\bfx}|f_2(k,t)|^2
\\ \label{corrdef}
&=& \frac{1}{2\pi^2r}\int_0^\infty dk ~k \sin(kr)|f_2(k,t)|^2
\pkt\eea
We have performed a series of simulations illustrating the transition from $N\to \infty$ to
finite $N$. Fig. 6a shows the correlation
functions $rC(r)$ at
times $t=30$, $50$ and $100$ for $\lambda \cdot N =1$, with $N=\infty$, $10$ and $4$.

\setcounter{fig2}{1}
\renewcommand{\thefigure}{\arabic{figure}\alph{fig2}}
\begin{figure}[htbp]
  \begin{center}
    \sffamily
    \label{fig:6a}
    \includegraphics[scale=0.4,angle=270]{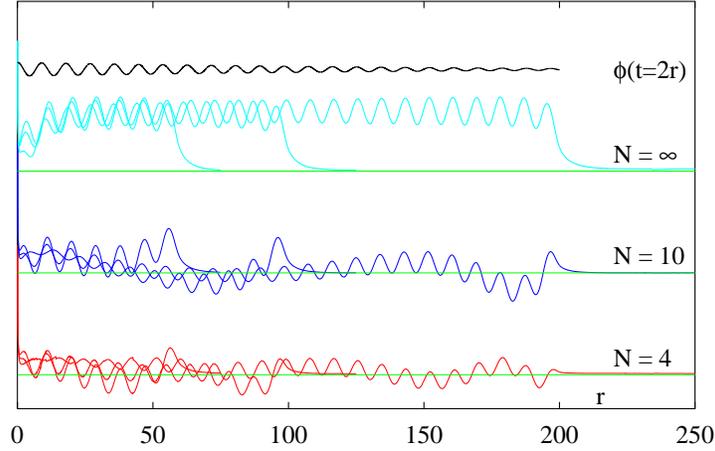}
    \caption{Equal-time correlations of pion fluctuations. We ßplot
$rC(r)$, with $C(r)$ as defined in Eq. \eqn{corrdef}, at times $t=30$,
$50$, $100$, for parameters $v=\sqrt{N}$, $\lambda=1/N$, $N=4$, $10$
and for $N\to \infty$. Initial amplitude $\phi_0=1.024 v$. On top
the amplitude $\phi(t)/v$ for $N \to \infty$.}
  \end{center}
\end{figure}
\addtocounter{figure}{-1}
\addtocounter{fig2}{1}

\begin{figure}[htbp]
  \begin{center}
    \sffamily
    \label{fig:6b}
    \includegraphics[scale=0.4,angle=270]{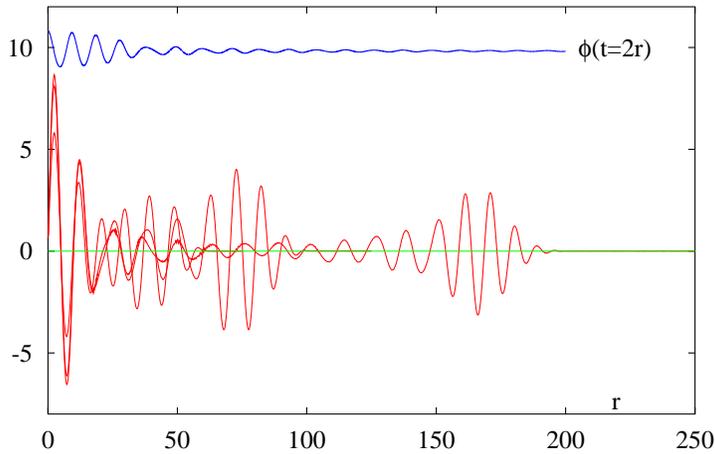}
    \caption{ Equal-time correlations of pion fluctuations; parameters
$N=4$, $\lambda=1/4$, $v=2$, initial amplitude $\phi_0=1.2 v$.}
  \end{center}
\end{figure}

The initial amplitude is $\phi_0ß\simeq 1.024 v$ implying in all cases $m_{20} \simeq 0.2$.
It is seen that there are long range correlations in all cases. They are
seemingly related to or actually generated by the oscillations of the classical field.
Its amplitude is displayed in the same figure for the case  $N\to \infty$, the
finite $N$ amplitudes behave very similarly. The correlations are positive
throughout for the large-$N$ case. For $N=10$ and $N=4$ they alternate in sign in the central
region at late times. So at small $N$, in particular at $N=4$,
these results certainly do not
suggest a growth of positively correlated domains but rather a wash-out
implied by the alternating signs. In fact the size of a positively correlated
region is essentially of the order of  an oscillation period of
the sigma field.

The close relationship between the oscillations of the classical field and
the correlations is further demonstrated in Fig. 6b.
For $N=4, \lambda=1$ and an initial value $\phi_0= 1.2 v $ the oscillations
of the classical field decay rather quickly, and so do the correlations.
This means that one has to be cautious with their interpretation: these correlations
are not due to an interaction between field fluctuations but are generated by the coherence
of such fluctuations with an external source. This is not necessarily the
wrong physics, but certainly this feature is inherent in the mean field
approach.

\subsection{\label{momspec}Momentum spectra}

 In the large-$N$ limit one of the most pronounced characteristic
of the momentum spectra of the ``pion'' fluctuations is the occurrence
of parametric resonance bands. These develop already in the early stage of
evolution, before backreaction sets in. Then the time dependence of the
 classical field  is described by Jacobian elliptic functions, and the
fluctuations are solutions of the Lam\'e equation. These solutions have been
derived and discussed extensively in \cite{Boyanovsky:1996sq,Boyanovsky:1998zg,
Boyanovsky:1999yp}.

For our finite-$N$ system the early time behavior of the classical field
and of the pion modes is the same as in the large-$N$ limit, as
before the onset of fluctuations both
$\calm_{ß\rm cl}^2$ and $\calm_2^2$ are approximatley equal to $\lambda (\phi^2-v^2)$, while
$\calm_1^2\simeq \lambda(3\phi^2-v^2)$. The analysis of the pion fluctuations can be taken
over from Ref. \cite{Boyanovsky:1996sq}, therefore.
For $\phi_0 < \sqrt{2} v$ the parametric resonance band is given by
\footnote{Here and in the following we simplify the discussion by neglecting
 modifications due to finite renormalization
corrections.}
\be \label{paramres1}
\lambda\left(v^2-\frac{\phi_0^2}{2}\right) < k^2 < \frac{\lambda}{2} \phi_0^2
\kma\ee
as given in Ref. \cite{Boyanovsky:1996sq}.
 For $\phi_0 > \sqrt{2} v$ parametric resonance occurs (see Appendix A)
in the momentum interval
\be \label{paramres2}
0 < k^2 < \frac{\lambda}{2} \phi_0^2\pkt
\ee

In Figs. 7 and 8 we plot the integrands of the various fluctuation integrals $\calf_i$, i.e.
\be \label{flintegs}
  \frac{k^2}{2\pi^2 2\omega_{i0}}\left(|f_i(k,t)|^2-1+\frac{\calv_i}{2\omega^2_{i0}}\right)
\ee
as functions of $k$.

\setcounter{fig2}{1}
\renewcommand{\thefigure}{\arabic{figure}\alph{fig2}}
\begin{figure}[htbp]
  \begin{center}
    \sffamily
    \label{fig:7a}
    \includegraphics[scale=0.4,angle=270]{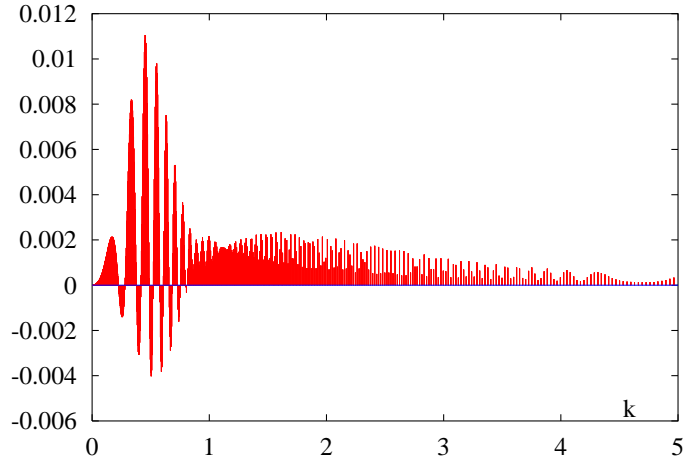}
    \caption{ Spectrum of sigma fluctuations in the broken symmetry phase.
We display the integrand defined in Eq. \eqn{flintegs} of the fluctuation integral
as a  function of $k$ for $\lambda=1$, $v=1$, $\phi_0=1.2 v$ at $t=100$.}
  \end{center}
\end{figure}
\addtocounter{figure}{-1}
\addtocounter{fig2}{1}

\begin{figure}[htbp]
  \begin{center}
    \sffamily
    \label{fig:7b}
    \includegraphics[scale=0.4,angle=270]{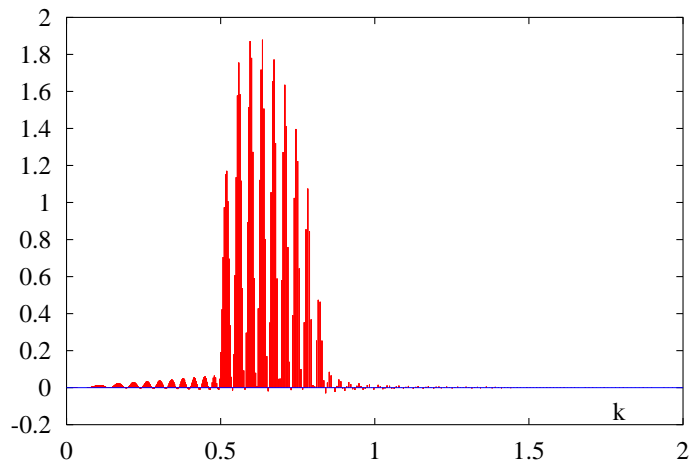}
    \caption{Spectrum of pion fluctuations in the broken symmetry phase.
Notation and parameters as in Fig. 7a.}
  \end{center}
\end{figure}

\setcounter{fig2}{1}
\renewcommand{\thefigure}{\arabic{figure}\alph{fig2}}
\begin{figure}[htbp]
  \begin{center}
    \sffamily
    \label{fig:8a}
    \includegraphics[scale=0.4,angle=270]{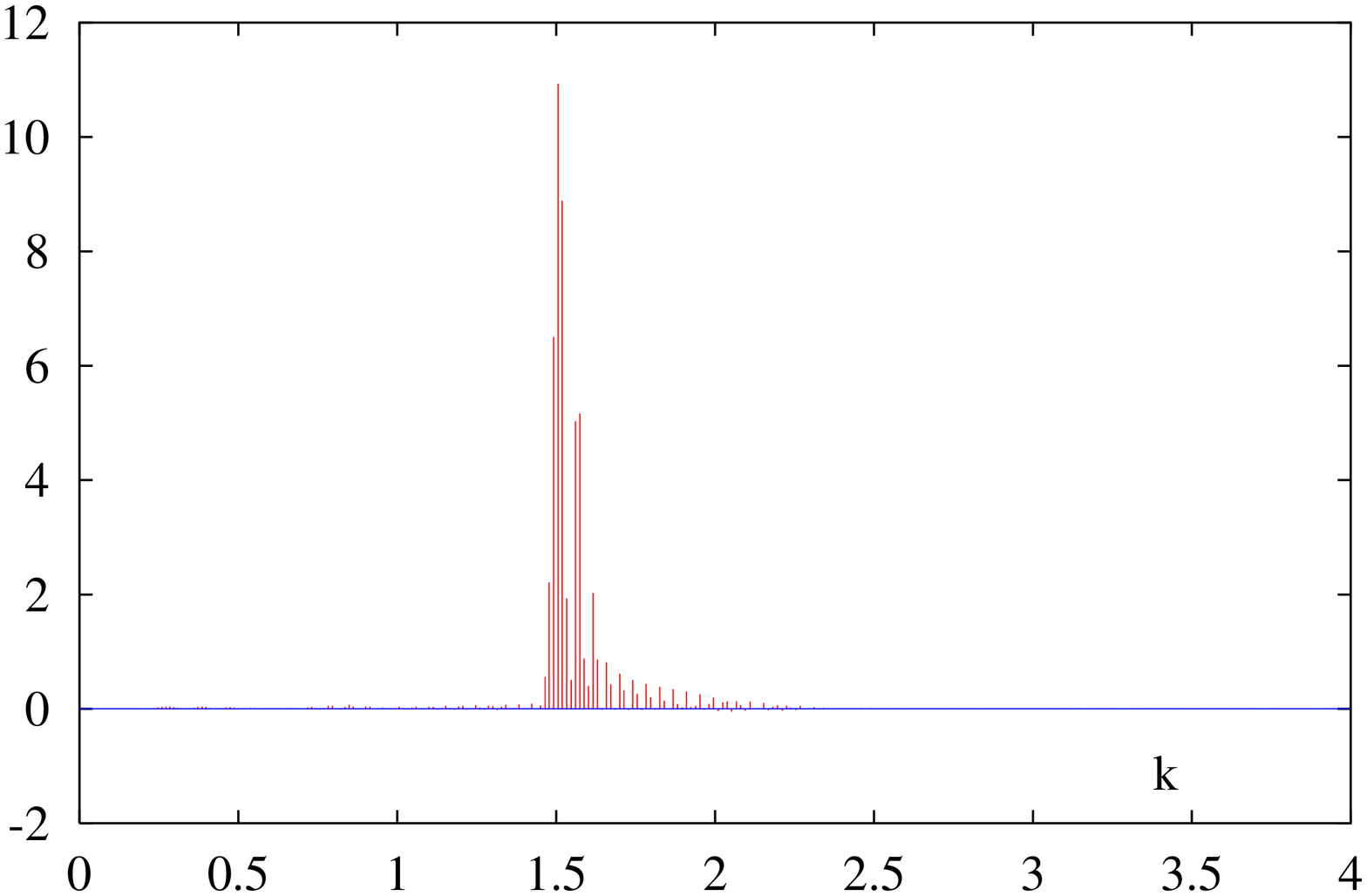}
    \caption{ Spectrum of sigma fluctuations in the symmetric phase.
Notation and parameters as in Fig. 7a, however $\phi_0=2v$.}
  \end{center}
\end{figure}
\addtocounter{figure}{-1}
\addtocounter{fig2}{1}

\begin{figure}[htbp]
  \begin{center}
    \sffamily
    \label{fig:8b}
    \includegraphics[scale=0.4,angle=270]{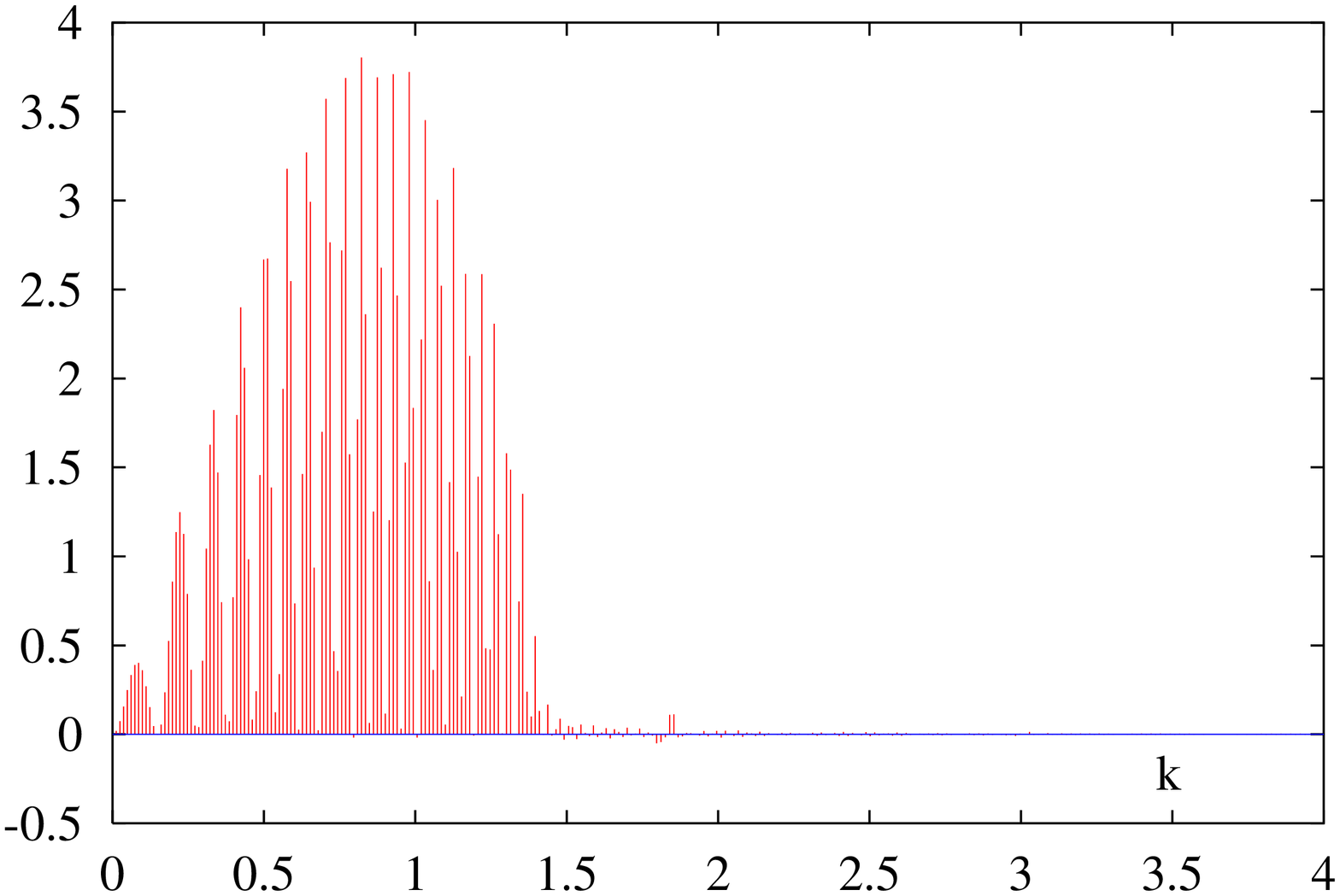}
    \caption{ Spectrum of pion fluctuations in the symmetric phase.
Notation and parameters as in Fig. 7a, however $\phi_0=2v$.}
  \end{center}
\end{figure}

As seen in Figs. 7b and 8b parametric resonance indeed dominates the pion fluctuations
within the momentum intervals \eqn{paramres1} and \eqn{paramres2}
in the broken symmetry and in the symmetric phase respectively.
The sigma fluctuations are unimportant in the broken symmetry phase,
as displayed in Fig. 7a.
In the symmetric phase they develop a pronounced resonance which seemingly
becomes sharper and sharper as time increases. The position of the resonance can be obtained
by considering the periodicity $T$ of the square of the classical field
and the frequency of the ``sigma'' fluctuations; one expects
$k^2+\langle\calm_1^2\rangle \simeq 4 \pi^ 2/T^ 2$. The empirical (``measured'') values
of $T$ and $\langle\calm_1^2\rangle$ for the simulation with our parameter
set yield $k\simeq ·1.5$ in agreement with Fig. 8a. As the resonance is due to a
time-dependent mass term, it should be situated in a resonance band, as indeed suggested
by Fig 8a. For this resonance band the early time analysis does not apply, as the sigma fluctuations
evolve at a time when backreaction due to pion fluctuations has already modified the
equation for the condensate field $\phi(t)$, see Fig. 2c. It should be noted that an analoguos
resonance appears in the spectrum of sigma (longitudinal) fluctuations in the $\Phi^ 4$ model
{\em without} spontaneous symmetry breaking, while the pion (transverse) fluctuations
display a broad parametric resonance band. So again the difference between the
fluctuations parallel and orthogonal to the classical field is not related to spontaneous
breaking but to the non-symmetry of the state.

\subsection{\label{dyneffpot}A naive dynamical effective potential}

 As we have seen, the behavior of the system below $\phi_0=\sqrt{2}v$ displays
the features of a spontaneously broken phase in a rather
convincing way, while for the symmetric
phase the non-symmetric initial conditions remain manifest even at late times.
The amplitude of oscillations of the classical field does not go to zero even
if we extend the simulation to times beyond which we believe our numerical
reliability.

\setcounter{fig2}{1}
\renewcommand{\thefigure}{\arabic{figure}\alph{fig2}}
\begin{figure}[htbp]
  \begin{center}
    \sffamily
    \label{fig:9a}
    \includegraphics[scale=0.4,angle=270]{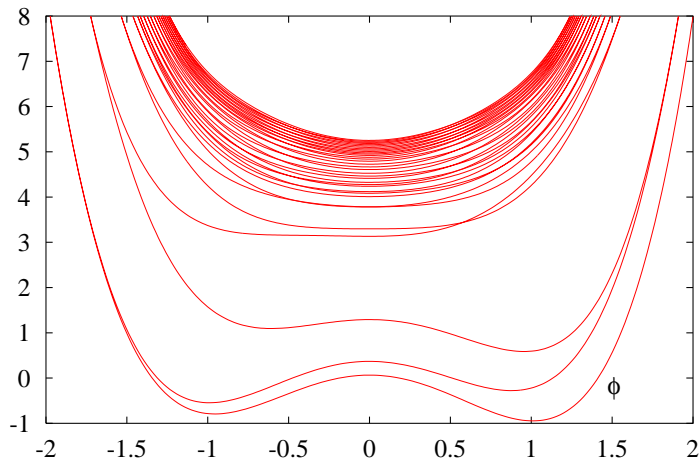}
    \caption{Evolution of the potential energy.
Definition as in subsection \ref{dyneffpot}. Parameters: $v=1$, $\lambda=1$,
and $\phi_0=1.2v$.}
  \end{center}
\end{figure}
\addtocounter{figure}{-1}
\addtocounter{fig2}{1}

\begin{figure}[htbp]
  \begin{center}
    \sffamily
    \label{fig:9b}
    \includegraphics[scale=0.4,angle=270]{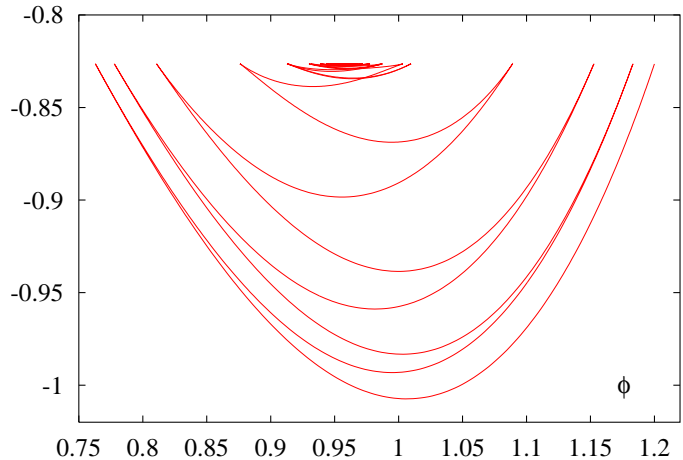}
    \caption{Evolution of the potential energy.
Parameters as in Fig. 9a, however $\phi_0=2v$.}
  \end{center}
\end{figure}

The evidence for symmetry restoration remains indirect, therefore. So it might be
useful to have yet another criterion. This is given by the dynamics of the classical field.
In a rather trivial way the potential that yields the ``force'' experienced by the
classical field is given by the potential energy, obtained by subtracting the kinetic
energy from the total energy, $V_{\rm pot}=E-\dot \phi^2/2$. This potential is of course
time-dependent and is explored only within the range of oscillations of the field.
We display this potential energy in Figs.  9a and 9b for the broken and for the symmetric
phase, respectively.
It is seen that the positions of minima  shift  from $\phi = \pm v$ to smaller
absolute values in the broken symmetry phase, while in the symmetric phase
the separate minima entirely disappear after a few oscillations
and a new minimum appears at $\phi=0$.

%%%%%%%%%%%%%%%%%%%%%%%%%%%%%%%%%%%%%%%%%%%%%%%%%%%%%%%%%%%%%%%%%%%%%%%%%%%%%%%%%%%%%%%%%%%%

\section{\label{conclus}Conclusions}
\setcounter{equation}{0}
We have presented here an analysis of nonequilibrium dynamics in $O(N)$ models
with spontaneous symmetry breaking
at finite $N$ in a bubble-resummed one-loop approximation.
This has allowed us to study some new features of such systems, not accessible
to the large-$N$ or one-loop approximations.

We have found that the back-reaction of the quantum fluctuations on themselves
prevents, as in the large-$N$ approximation, a catastrophic instability
of the system as found in the one-loop approximation. The mechanism is, like
in large-$N$, an exponential evolution of those quantum fluctuations whose
effective mass squared has become negative, pushing this mass squared
back to positive values.

The dependence on the initial conditions displays features of an expected
phase transition between a regime with spontaneous symmetry breaking and
a symmetric phase. A more detailed analysis of nonequilibrium dynamics in the critical region
may provide new insights into the  nature of the phase transition.

Our approach provides an new tool for performing studies of nonequilibrium evolution
in scenarios typical of heavy ion collisions \cite{Muller:1999wa}
or of the dynamics of inflation in the early universe
\cite{Linde:2000kn}, using appropriate models
like the $O(4)$ sigma model or realistic grand unified theories.
It should be especially useful for studying phase transitions
and critical behavior in an nonequilibrium
context. It would be useful to extend this study to more general initial conditions
exploiting more of the multi-dimensional mexican hat structure of the potentials
in models with $N\neq1$. The techniques using coupled-channel Green functions
are available \cite{Baacke:1990zu,Baacke:1997hw,Cormier:2001iw}.

Thermalization is not expected in the large-$N$ approximation and, in the absence of a rescattering
involving the next order sunset diagrams, may not be expected here either. One may consider this as
a drawback of the Hartree type approximation we are using. In simpler quantum systems
where one can compare to the exact time evolution  the
Hartree approximation is found to constitute an improvement with respect
the large-$N$ approximation, but is further improved by the inclusion of sunset diagrams
\cite{Mihaila:2000ib,Mihaila:2001sr,Blagoev:2001ze}.
Still it describes well the early time behavior.
In heavy ion collisions or in
the early universe the behavior at early
times, and/or without reaching equilibrium and thermalization, may be even more realistic
and is therefore interesting as well.  It is rewarding
that the phase structure of the theory is revealed  by nonequilibrium dynamics at early times
already.

Thermalization is, on the other hand, a basic theoretical issue, independent of
realization in concrete physical processes.
It has recently been investigated by studying the time evolution
of classical fluctuations in Hartree approximation \cite{Aarts:2000mg}.
As the present formalism can be extended to include higher-loop diagrams
\cite{Verschelde:2001dz,Verschelde:2000ta,Smet:2001un}  our work can
be considered as a first step towards calculations
including rescattering of fluctuations \cite{Berges:2000ur} and
of ``controlled nonperturbative
dynamics'' \cite{Berges:2001fi,Aarts:2001qa} in three dimensions, using a continuum regularization.
This requires the inclusion of sunset and higher order diagrams, incorporating
rescattering of the quantum fluctuations.
Such calculations are being performed at present
\cite{Bergespc}, using lattice regularizations.
We think that it will still take a long time until the limitations introduced
by various approximations are fully understood, leading
to a well-based understanding of thermalization in quantum field theory.
So various alternative approaches will have to be considered.
With our modest new step, introducing {\em some} $1/N$ corrections
we are clearly still far off such a
demanding formal and in particular numerical task; we think, however, that
our investigation provides some useful, and possibly inspiring,
new insights.

\section*{Acknowledgments}
The authors have pleasure in thanking G. Aarts, J. Berges and A. Heinen for
useful and stimulating discussions.

\renewcommand{\theequation}{\Alph{section}.\arabic{equation}}
\begin{appendix}
\setcounter{equation}{0}

\section{Some solutions of the Lam\'e equation}
The following analysis follows closely the one in Ref.
\cite{Boyanovsky:1996sq}. The basic formulae are from Ref. \cite{Abramowicz},
the only difference in notation is in the argument of the elliptic integral
$K$ which is $m$ there and $k=\sqrt m$ here.
We introduce the dimensionless classical field  $\chi=\phi/v$ and the dimensionless
time variable $\tau=\sqrt \lambda v t$. We denote differentiation with respect to
$\tau$  by a prime, $df/d\tau=f'$. Then the equation of motion for $\chi$ before
the onset of backreaction via fluctuations, and neglecting finite renormalization terms
\footnote{This simplification is not really necessary, but appropriate for
the couplings used here.} is given by
\be
\chi'' +\chi^ 3 - \chi =0
\pkt
\ee
Using the dimensionless momentum variable $q=k/\sqrt \lambda v$ (here $k$ is the
momentum variable used in the main text) the mode equations
before the onset of backreaction read
\bea
f_1'' + (q^ 2 -1 +3 \chi^ 2)f_1&=&0
\\
f_2'' + (q^ 2 -1 + \chi^ 2)f_2 &=& 0
\pkt\eea
For the symmetric phase, i.e., $\chi_0=\phi_0/v > \sqrt 2$, the solution
of the classical equation of motion is given by
\be
\chi(\tau)=\chi_0\left[1-{\rm sn}^ 2(\tau \sqrt{\chi_0^ 2-1},k)\right]^{1/2}
\ee
where ${\rm sn}$ denotes the Jacobi elliptic function, its index $k$
is given by $k=\chi_0/\sqrt{2(\chi_0^2-1)}$. It may be related to a 
Weierstra\ss\ elliptic function $\calp$ via
\be
{\rm sn}^ 2(\tau \sqrt{\chi_0^ 2-1},k)=\frac{1}{k^2(e1-e3)}
\left[\calp\left(\frac{\tau \sqrt{\chi_0^ 2-1}+iK'(k)´}{\sqrt{e_1-e_3}}\right)-e_3\right]
\pkt\ee
Furthermore we have (see section 18.9 of Ref. \cite{Abramowicz})
$i K'(k)=\omega'\sqrt{e_1-e_3}$ where $\omega'$ is the imaginary half period
of the Weierstra\ss\ function $\calp$.
For the roots
 $e_i$ we find $e_1=\chi_0^ 2/2-2/3$, $e_2=1/3$, and
$e_3=-\chi_0^ 2/2+1/3$, the invariants are given by
$g_2=(\chi_0^2-1)^2+1/3$ and $3g_3=-(\chi_0^2-1)^2+1/9$.
The half periods of the double-periodic function $\calp$ are related to the roots by
$\calp(\omega)=e_1$, $\calp(\omega+\omega')=e_2$ and $\calp(\omega')=e_3$.
The equations of motion for the $f_i$ then become
\bea
f_1'' +\left[q^2+1 -6\calp(\tau+\omega'´) \right]f_1 &=& 0
\\
f_2''+ \left[ q^ 2 -\frac{1}{3} -2 \calp(\tau + \omega'´)\right]f_2 &=& 0
\pkt\eea
The general Lam\'e equation reads \cite{Kamke}
\be
f'' -[a +n(n+1)\calp]f =0
\pkt \ee
For $f_2$ we have $n=1$ and the solution is given by
\be
f_2 =\frac{\sigma(\tau\pm\alpha)}{\sigma(\tau)}\exp(\mp \tau \zeta(\alpha))
\kma\ee
where $\alpha$ is determined by the equation
\be \label{realax2}
\calp(\alpha)=a
\kma
\ee and where
$\sigma$ and $\zeta$ are Weierstra\ss\ functions.

For $f_1$ we have $n=2$ and the solution is given by \cite{Kamke}
\be
f_1=\frac{d}{d\tau}\frac{\sigma(\tau\pm\alpha)}{\sigma(\tau)}\exp(\mp\tau(\zeta(\alpha)+\beta))
\ee
where $\alpha$ is one of the solutions of
\be \label{realax1}
\calp(\alpha)=\frac{a_1^3+g_3}{3a_1^2-g_2}
\ee
with $a_1=a/3$, and where
\be
\beta=\frac{\calp'(\alpha)}{2 \calp(\alpha)-a_1}
\pkt\ee

The solutions of the Lam\' e equation are quasiperiodic; if the argument increases
by a period $2\omega$ then the solution reproduces itself up to factor
$\exp (iF(\alpha))$ where $F(\alpha)$ is known as Floquet index. If it is imaginary
then there is an exponentially increasing solution, if it is real the
solutions are periodic up to a phase.
The right hand sides of \eqn{realax1} and \eqn{realax2} are real. The 
Weierstra\ss\
function $\calp$ maps the fundamental rectangle $[0,\omega,\omega+\omega',\omega',0]$
onto the upper half plane, so the solutions $\alpha$ of these equations are situated
on the boundary of the fundamental rectangle; the origin $\alpha=0$ is mapped to
the infinite point.

For $f_2$ a closer analysis shows that on the
sections  $[0,\omega]$ and $[\omega',\omega'+\omega] $ of this boundary
the Floquet index
\be
F(\alpha)=-2i\left[\alpha \zeta(\omega)-\omega \zeta(\alpha)\right]
\ee
is imaginary, so in these regions one has
parametric resonance (``forbidden bands'' in analogy to solutions of the 
Schr\" odinger
equation in periodic potentials);
on the sections  $[0,\omega']$ and $[\omega,\omega+\omega']$
the solutions are oscillatory (``allowed bands'').
The ``forbidden'' bands are characterized by
$e_1 < \calp(\alpha) < \infty$ and $e_3 < \calp(\alpha) < e_2$.
For $f_2$ this implies parametric resonance in the in the intervals
$-\infty < q^ 2 <-\chi_0^ 2/2+1<0$ which is excluded kinematically, and in the
interval $0 < q ^2 < \chi_0^ 2/2$. This  resonance band manifests
itself in Fig. $8b$.

For $f_1$, the sigma fluctuations, the analysis is somewhat more cumbersome,
Eq. \eqn{realax1} becomes explicitly
\be \label{realax11}
\calp(\alpha)=-\frac{1}{9}~\frac{(1+q^ 2)^ 3-1+9(\chi_0^ 2-1)^ 2}{(1+q^ 2)^2-1-3(\chi_0^2-1)^2}
\pkt\ee
Again $\alpha$ is situated on the boundary of the fundamental rectangle with the same
corners.
The Floquet index is found to be
\be
i F(\alpha)=\alpha \zeta(\omega)-\omega \zeta(\alpha)
-\omega \frac{\calp'(\alpha)}{2\calp(\alpha)+(1+q^2)/3}
\pkt
\ee
It is imaginary for values of $\alpha$ on the sections
$[0,\omega]$ and $[\omega',\omega'+\omega] $ of the boundary of the fundamental
rectangle. Again this maps onto the intervals $e_1 < \calp(\alpha) < \infty$ and
$e_3 < \calp(\alpha) < e_2$.
The  analysis of the function on the right hand side of Eq. \eqn{realax11}
shows that the resonance bands are $-\infty < q^ 2 <-1 -\sqrt{3(\chi_0^2-1)^2+1}$ and
$-3\chi_0^2/2 < q^ 2 < 0$ which are excluded
kinematically, and, in the physical region,
 $3 \chi_0^ 2/2-3 < q^2 < -1+\sqrt{3(\chi_0^ 2-1)^2+1}$.
The latter resonance band is {\em not} manifest in Fig. 8a as the sigma fluctuations
develop only after the classical equation of motion is modified by the backreaction
due to the pion fluctuations. We have verified that our analytical result
for the resonance band  is correct
by running the simulation without backreaction. The parametric resonance of the sigma fluctuations
analyzed here may manifest itself for other parameter sets and the result
may be of importance, therefore.

\end{appendix}

\newpage


\begin{thebibliography}{99}

%\cite{Coleman:1974jh}
\bibitem{Coleman:1974jh}
S.~Coleman, R.~Jackiw and H.~D.~Politzer,
%``Spontaneous Symmetry Breaking In The O(N) Model For Large N*,''
Phys.\ Rev.\ D {\bf 10}, 2491 (1974).
%%CITATION = PHRVA,D10,2491;%%

%\cite{Dolan:1974qd}
\bibitem{Dolan:1974qd}
L.~Dolan and R.~Jackiw,
%``Symmetry Behavior At Finite Temperature,''
Phys.\ Rev.\ D {\bf 9}, 3320 (1974).
%%CITATION = PHRVA,D9,3320;%%



%\cite{Bardeen:1983st}
\bibitem{BarMosh}
W.~A.~Bardeen and M.~Moshe,
%``Phase Structure Of The O(N) Vector Model,''
Phys.\ Rev.\ D {\bf 28}, 1372 (1983);
%%CITATION = PHRVA,D28,1372;%%
%\cite{Bardeen:1986td}
%W.~A.~Bardeen and M.~Moshe,
%``Comment On The Finite Temperature Behavior Of Lambda (Phi**2)**2 In Four-Dimensions Theory,''
Phys.\ Rev.\ D {\bf 34}, 1229 (1986).
%%CITATION = PHRVA,D34,1229;%%


%\cite{Calzetta:1987ey}
\bibitem{Calzetta:1987ey}
E.~Calzetta and B.~L.~Hu,
%``Closed Time Path Functional Formalism In Curved Space-Time: Application To Cosmological Back Reaction Problems,''
Phys.\ Rev.\ D {\bf 35}, 495 (1987);
%%CITATION = PHRVA,D35,495;%%
%\cite{Calzetta:1988cq}
%\bibitem{Calzetta:1988cq}
%E.~Calzetta and B.~L.~Hu,
%%``Nonequilibrium Quantum Fields: Closed Time Path Effective Action, Wigner Function And Boltzmann Equation,''
Phys.\ Rev.\ D {\bf 37}, 2878 (1988).
%%CITATION = PHRVA,D37,2878;%%

%
%\cite{Cooper:1994hr}
\bibitem{Cooper:1994hr}
F.~Cooper, S.~Habib, Y.~Kluger, E.~Mottola, J.~P.~Paz and P.~R.~Anderson,
%``Nonequilibrium quantum fields in the large N expansion,''
Phys.\ Rev.\ D {\bf 50}, 2848 (1994)
[arXiv:hep-ph/9405352].
%%CITATION = HEP-PH 9405352;%%

%\cite{Boyanovsky:1994xf}
\bibitem{Boyanovsky:1994xf}
D.~Boyanovsky, H.~J.~de Vega and R.~Holman,
%``Nonequilibrium evolution of scalar fields in FRW cosmologies I,''
Phys.\ Rev.\ D {\bf 49}, 2769 (1994)
[arXiv:hep-ph/9310319].
%%CITATION = HEP-PH 9310319;%%

%\cite{Boyanovsky:1995me}
\bibitem{Boyanovsky:1995me}
D.~Boyanovsky, H.~J.~de Vega, R.~Holman, D.~S.~Lee and A.~Singh,
%``Dissipation via particle production in scalar field theories,''
Phys.\ Rev.\ D {\bf 51}, 4419 (1995)
[arXiv:hep-ph/9408214].
%%CITATION = HEP-PH 9408214;%%

%\cite{Boyanovsky:1995em}
\bibitem{Boyanovsky:1995em}
D.~Boyanovsky, M.~D'Attanasio, H.~J.~de Vega, R.~Holman and D.~-.~Lee,
%``Reheating and thermalization: Linear versus nonlinear relaxation,''
Phys.\ Rev.\ D {\bf 52}, 6805 (1995)
[arXiv:hep-ph/9507414].
%%CITATION = HEP-PH 9507414;%%

%\cite{Baacke:1997se}
\bibitem{Baacke:1997se}
J.~Baacke, K.~Heitmann and C.~Patzold,
%``Nonequilibrium dynamics: A renormalized computation scheme,''
Phys.\ Rev.\ D {\bf 55}, 2320 (1997)
[arXiv:hep-th/9608006].
%%CITATION = HEP-TH 9608006;%%

%\cite{Cooper:1997ii}
\bibitem{Cooper:1997ii}
F.~Cooper, S.~Habib, Y.~Kluger and E.~Mottola,
%``Nonequilibrium dynamics of symmetry breaking in lambda Phi**4 field  theory,''
Phys.\ Rev.\ D {\bf 55}, 6471 (1997)
[arXiv:hep-ph/9610345].
%%CITATION = HEP-PH 9610345;%%

%\cite{Boyanovsky:1996sq}
\bibitem{Boyanovsky:1996sq}
D.~Boyanovsky, H.~J.~de Vega, R.~Holman and J.~F.~Salgado,
%``Analytic and numerical study of preheating dynamics,''
Phys.\ Rev.\ D {\bf 54}, 7570 (1996)
[arXiv:hep-ph/9608205].
%%CITATION = HEP-PH 9608205;%%

%\cite{Boyanovsky:1998zg}
\bibitem{Boyanovsky:1998zg}
D.~Boyanovsky, C.~Destri, H.~J.~de Vega, R.~Holman and J.~Salgado,
%``Asymptotic dynamics in scalar field theory: Anomalous relaxation,''
Phys.\ Rev.\ D {\bf 57}, 7388 (1998)
[arXiv:hep-ph/9711384].
%%CITATION = HEP-PH 9711384;%%

%\cite{Boyanovsky:1999yp}
\bibitem{Boyanovsky:1999yp}
D.~Boyanovsky, H.~J.~de Vega, R.~Holman and J.~Salgado,
%``Non-equilibrium Bose-Einstein condensates, dynamical scaling and  symmetric evolution in large N phi**4 theory,''
Phys.\ Rev.\ D {\bf 59}, 125009 (1999)
[arXiv:hep-ph/9811273].
%%CITATION = HEP-PH 9811273;%%

%\cite{Baacke:2000fw}
\bibitem{Baacke:2000fw}
J.~Baacke and K.~Heitmann,
%``Nonequilibrium evolution and symmetry structure of the large-N phi**4  model at finite temperature,''
Phys.\ Rev.\ D {\bf 62}, 105022 (2000)
[arXiv:hep-ph/0003317].
%%CITATION = HEP-PH 0003317;%%




%\cite{Lenaghan:2000si}
\bibitem{Lenaghan:2000si}
J.~T.~Lenaghan and D.~H.~Rischke,
%``The O(N) model at finite temperature: Renormalization of the gap  equations in Hartree and large-N approximation,''
J.\ Phys.\ G {\bf G26}, 431 (2000)
[nucl-th/9901049].
%%CITATION = NUCL-TH 9901049;%%

%\cite{Destri:2000hd}
%\bibitem{Destri:2000hd}
\bibitem{DM}
C.~Destri and E.~Manfredini,
%``Out-of-equilibrium dynamics of large-N phi**4 QFT in finite volume,''
Phys.\ Rev.\ D {\bf 62}, 025007 (2000)
[arXiv:hep-ph/0001177];
%%CITATION = HEP-PH 0001177;%%
%\cite{Destri:2000he}
%\bibitem{Destri:2000he}
%C.~Destri and E.~Manfredini,
%``An improved time-dependent Hartree-Fock approach for scalar phi**4 QFT,''
Phys.\ Rev.\ D {\bf 62}, 025008 (2000)
[arXiv:hep-ph/0001178].
%%CITATION = HEP-PH 0001178;%%


%\cite{Amelino-Camelia:1993nc}
\bibitem{Amelino-Camelia:1993nc}
G.~Amelino-Camelia and S.~Pi,
%``Selfconsistent improvement of the finite temperature effective potential,''
Phys.\ Rev.\ D {\bf 47}, 2356 (1993)
[arXiv:hep-ph/9211211].
%%CITATION = HEP-PH 9211211;%%

%\cite{Amelino-Camelia:1997dd}
\bibitem{Amelino-Camelia:1997dd}
G.~Amelino-Camelia,
%``Thermal effective potential of the O(N) linear sigma model,''
Phys.\ Lett.\ B {\bf 407}, 268 (1997)
[arXiv:hep-ph/9702403].
%%CITATION = HEP-PH 9702403;%%

%\cite{Chiku:1998va}
%\bibitem{Chiku:1998va}
\bibitem{ChiHa}
S.~Chiku and T.~Hatsuda,
%``Soft modes associated with chiral transition at finite temperature,''
Phys.\ Rev.\ D {\bf 57}, 6 (1998)
[arXiv:hep-ph/9706453];
%%CITATION = HEP-PH 9706453;%%
%\cite{Chiku:1998kd}
%\bibitem{Chiku:1998kd}
%S.~Chiku and T.~Hatsuda,
%``Optimized perturbation theory at finite temperature,''
Phys.\ Rev.\ D {\bf 58}, 076001 (1998)
[arXiv:hep-ph/9803226].
%%CITATION = HEP-PH 9803226;%%

%\cite{Cornwall:1974vz}
\bibitem{Cornwall:1974vz}
J.~M.~Cornwall, R.~Jackiw and E.~Tomboulis,
%``Effective Action For Composite Operators,''
Phys.\ Rev.\ D {\bf 10}, 2428 (1974).
%%CITATION = PHRVA,D10,2428;%%

%\cite{Nemoto:2000qf}
\bibitem{Nemoto:2000qf}
Y.~Nemoto, K.~Naito and M.~Oka,
%``Effective potential of O(N) linear sigma model at finite temperature,''
Eur.\ Phys.\ J.\ A {\bf 9}, 245 (2000)
[arXiv:hep-ph/9911431].
%%CITATION = HEP-PH 9911431;%%
%\cite{Verschelde:2001dz}

\bibitem{Verschelde:2001dz}
H.~Verschelde,
%``Summation and renormalization of bubble graphs to all orders,''
Phys.\ Lett.\ B {\bf 497}, 165 (2001)
[arXiv:hep-th/0009123].
%%CITATION = HEP-TH 0009123;%%

%\cite{Verschelde:2000ta}
\bibitem{Verschelde:2000ta}
H.~Verschelde and J.~De Pessemier,
%``Study of the O(N) linear sigma model at finite temperature using the  2PPI expansion,''
arXiv:hep-th/0009241.
%%CITATION = HEP-TH 0009241;%%

%\cite{Smet:2001un}
\bibitem{Smet:2001un}
G.~Smet, T.~Vanzielighem, K.~Van Acoleyen and H.~Verschelde,
%``A 2 loop 2PPI analysis of lambda phi**4 at finite temperature,''
arXiv:hep-th/0108163.
%%CITATION = HEP-TH 0108163;%%

%\cite{Muller:1999wa}
\bibitem{Muller:1999wa}
see, e.g.: B.~.~Muller and R.~D.~Pisarski,
`RHIC physics and beyond: Kay Kay Gee Day. Proceedings, Upton, Brookhaven, USA, October 23, 1998,''
{\it  Woodbury, USA: AIP (1999) 167 p}.

%\cite{Gavin:1993px}
\bibitem{Gavin:1993px}
S.~Gavin and B.~Muller,
%``Larger domains of disoriented chiral condensate through annealing,''
Phys.\ Lett.\ B {\bf 329}, 486 (1994)
[arXiv:hep-ph/9312349].
%%CITATION = HEP-PH 9312349;%%

%\cite{Mrowczynski:1995at}
\bibitem{Mrowczynski:1995at}
S.~Mrowczynski and B.~Muller,
%``Reheating after supercooling in the chiral phase transition,''
Phys.\ Lett.\ B {\bf 363}, 1 (1995)
[arXiv:nucl-th/9507033].
%%CITATION = NUCL-TH 9507033;%%

%\cite{Kaiser:1998hf}
\bibitem{Kaiser:1998hf}
D.~I.~Kaiser,
%``Larger domains from resonant decay of disordered chiral condensates,''
Phys.\ Rev.\ D {\bf 59}, 117901 (1999)
[arXiv:hep-ph/9801307].
%%CITATION = HEP-PH 9801307;%%

%\cite{Hiro-Oka:1999xk}
\bibitem{Hiro-Oka:1999xk}
H.~Hiro-Oka and H.~Minakata,
%``Dynamical pion production via parametric resonance from disoriented  chiral condensate,''
Phys.\ Rev.\ C {\bf 61}, 044903 (2000)
[arXiv:hep-ph/9906301];
%%CITATION = HEP-PH 9906301;%%
%\cite{Hiro-Oka:2001hx}
%\bibitem{Hiro-Oka:2001hx}
%H.~Hiro-Oka and H.~Minakata,
%``Pion production by parametric resonance mechanism with quantum back  reactions,''
Phys.\ Rev.\ C {\bf 64}, 044902 (2001)
[arXiv:hep-ph/0103181].
%%CITATION = HEP-PH 0103181;%%

%\cite{GomezNicola:2001js}
\bibitem{GomezNicola:2001js}
A.~Gomez Nicola,
%``Pion production in nonequilibrium chiral perturbation theory,''
Phys.\ Rev.\ D {\bf 64}, 016011 (2001)
[arXiv:hep-ph/0103198].
%%CITATION = HEP-PH 0103198;%%

%\cite{Linde:2000kn}
\bibitem{Linde:2000kn}
see, e.g.: A.~D.~Linde,
%``Inflationary cosmology,''
Phys.\ Rept.\  {\bf 333} (2000) 575.
%%CITATION = PRPLC,333,575;%%

%\cite{Dolgov:1982th}
\bibitem{Dolgov:1982th}
A.~D.~Dolgov and A.~D.~Linde,
%``Baryon Asymmetry In Inflationary Universe,''
Phys.\ Lett.\ B {\bf 116}, 329 (1982).
%%CITATION = PHLTA,B116,329;%%

%\cite{Abbott:1982hn}
\bibitem{Abbott:1982hn}
L.~F.~Abbott, E.~Farhi and M.~B.~Wise,
%``Particle Production In The New Inflationary Cosmology,''
Phys.\ Lett.\ B {\bf 117}, 29 (1982).
%%CITATION = PHLTA,B117,29;%%

%\cite{Kofman:1994rk}
\bibitem{Kofman:1994rk}
L.~Kofman, A.~D.~Linde and A.~A.~Starobinsky,
%``Reheating after inflation,''
Phys.\ Rev.\ Lett.\  {\bf 73}, 3195 (1994)
[arXiv:hep-th/9405187].
%%CITATION = HEP-TH 9405187;%%


%\cite{Boyanovsky:1996rw}
\bibitem{Boyanovsky:1996rw}
D.~Boyanovsky, D.~Cormier, H.~J.~de Vega and R.~Holman,
%``Out of equilibrium dynamics of an inflationary phase transition,''
Phys.\ Rev.\ D {\bf 55}, 3373 (1997)
[arXiv:hep-ph/9610396].
%%CITATION = HEP-PH 961039

%\cite{Ramsey:1997sa}
\bibitem{Ramsey:1997sa}
S.~A.~Ramsey and B.~L.~Hu,
%``Nonequilibrium inflaton dynamics and reheating: Back reaction of  parametric particle creation and curved spacetime effects,''
Phys.\ Rev.\ D {\bf 56}, 678 (1997)
[Erratum-ibid.\ D {\bf 57}, 3798 (1997)]
[arXiv:hep-ph/9706207];
%%CITATION = HEP-PH 9706207;%%
%\cite{Ramsey:1997qc}
%\bibitem{Ramsey:1997qc}
%S.~A.~Ramsey and B.~L.~Hu,
%``O(N) quantum fields in curved spacetime,''
Phys.\ Rev.\ D {\bf 56}, 661 (1997)
[arXiv:gr-qc/9706001].
%%CITATION = GR-QC 9706001;%%

%\cite{Shtanov:1994ce}
\bibitem{Shtanov:1994ce}
Y.~Shtanov, J.~Traschen and R.~H.~Brandenberger,
%``Universe reheating after inflation,''
Phys.\ Rev.\ D {\bf 51}, 5438 (1995)
[arXiv:hep-ph/9407247].
%%CITATION = HEP-PH 9407247;%%




%\cite{Mihaila:2000ib}
\bibitem{Mihaila:2000ib}
B.~Mihaila, T.~Athan, F.~Cooper, J.~Dawson and S.~Habib,
%``Exact and approximate dynamics of the quantum mechanical O(N) model,''
Phys.\ Rev.\ D {\bf 62}, 125015 (2000)
[arXiv:hep-ph/0003105].
%%CITATION = HEP-PH 0003105;%%

%\cite{Mihaila:2001sr}
\bibitem{Mihaila:2001sr}
B.~Mihaila, F.~Cooper and J.~F.~Dawson,
%``Resumming the large-N approximation for time evolving quantum systems,''
Phys.\ Rev.\ D {\bf 63}, 096003 (2001)
[arXiv:hep-ph/0006254].
%%CITATION = HEP-PH 0006254;%%

%\cite{Blagoev:2001ze}
\bibitem{Blagoev:2001ze}
K.~Blagoev, F.~Cooper, J.~Dawson and B.~Mihaila,
%``Schwinger-Dyson approach to non-equilibrium classical field theory,''
arXiv:hep-ph/0106195.
%%CITATION = HEP-PH 0106195;%%

%\cite{Aarts:2000mg}
\bibitem{Aarts:2000mg}
G.~Aarts, G.~F.~Bonini and C.~Wetterich,
%``On thermalization in classical scalar field theory,''
Nucl.\ Phys.\ B {\bf 587}, 403 (2000)
[arXiv:hep-ph/0003262].
%%CITATION = HEP-PH 0003262;%%

%\cite{Berges:2000ur}
\bibitem{Berges:2000ur}
J.~Berges and J.~Cox,
%``Thermalization of quantum fields from time-reversal invariant evolution  equations,''
Phys.\ Lett.\ B {\bf 517}, 369 (2001)
[arXiv:hep-ph/0006160].
%%CITATION = HEP-PH 0006160;%%

%\cite{Berges:2001fi}
\bibitem{Berges:2001fi}
J.~Berges,
%``Controlled nonperturbative dynamics of quantum fields out of  equilibrium,''
arXiv:hep-ph/0105311.
%%CITATION = HEP-PH 0105311;%%

%\cite{Aarts:2001qa}
\bibitem{Aarts:2001qa}
G.~Aarts and J.~Berges,
%``Nonequilibrium time evolution of the spectral function in quantum field  theory,''
arXiv:hep-ph/0103049.
%%CITATION = HEP-PH 0103049;%%



%\cite{Baacke:1998zy}
\bibitem{Baacke:1998zy}
J.~Baacke, K.~Heitmann and C.~Patzold,
%``Renormalization of nonequilibrium dynamics at large N and finite  temperature,''
Phys.\ Rev.\ D {\bf 57}, 6406 (1998)
[arXiv:hep-ph/9712506].
%%CITATION = HEP-PH 9712506;%%


%\cite{Cooper:1987pt}
\bibitem{Cooper:1987pt}
F.~Cooper and E.~Mottola,
%``Initial Value Problems In Quantum Field Theory In The Large N Approximation,''
Phys.\ Rev.\ D {\bf 36}, 3114 (1987).
%%CITATION = PHRVA,D36,3114;%%

%\cite{Maslov:1998bf}
\bibitem{Maslov:1998bf}
V.~P.~Maslov and O.~Y.~Shvedov,
%``Initial conditions in quasi-classical field theory,''
Theor.\ Math.\ Phys.\  {\bf 114}, 184 (1998)
[Teor.\ Mat.\ Fiz.\  {\bf 114}, 233 (1998)]
[arXiv:hep-th/9709151].
%%CITATION = HEP-TH 9709151;%%

%\cite{Baacke:1998zz}
\bibitem{Baacke:1998zz}
J.~Baacke, K.~Heitmann and C.~Patzold,
%``On the choice of initial states in nonequilibrium dynamics,''
Phys.\ Rev.\ D {\bf 57}, 6398 (1998)
[arXiv:hep-th/9711144].
%%CITATION = HEP-TH 9711144;%%

%\cite{Baacke:1990zu}
\bibitem{Baacke:1990zu}
J.~Baacke,
%``The Effective Action Of A Spin 1/2 Field In The Background Of A Nontopological Soliton,''
Z.\ Phys.\ C {\bf 47}, 619 (1990).
%%CITATION = ZEPYA,C47,619;%%

%\cite{Baacke:1997hw}
\bibitem{Baacke:1997hw}
J.~Baacke and A.~Surig,
%``Computing numerically the functional derivative of an effective action,''
Z.\ Phys.\ C {\bf 73}, 369 (1997)
[arXiv:hep-ph/9511231].
%%CITATION = HEP-PH 9511231;%%


%\cite{Cormier:2001iw}
\bibitem{Cormier:2001iw}
D.~Cormier, K.~Heitmann and A.~Mazumdar,
%``Dynamics of coupled bosonic systems with applications to preheating,''
arXiv:hep-ph/0105236.
%%CITATION = HEP-PH 0105236;%%


\bibitem{Bergespc}
J. Berges, private communication.


\bibitem{Ba_Mi_inprep}
J. Baacke and S. Michalski, in preparation


\bibitem{Abramowicz}
M. Abramowicz and I. A. Stegun (eds.), {\em Handbook
of Mathematical functions}, Dover Pubications, ,
New York 1965.

\bibitem{Kamke}
Erich Kamke, {\em Differentialgleichungen}, Vol. 1,
Akademische Verlagsgesellschaft, Leipzig 1967.


\end{thebibliography}
\end{document}